\documentclass[12pt,preprint]{aastex}
\usepackage{lineno}
\usepackage{multirow}
\usepackage{rotating}

\shorttitle{Long-term Observations at TeV Energies of HESS J0632+057}
\shortauthors{VERITAS and H.E.S.S. Collaborations}

\def \deg{^\circ}
\def \arcmin {\hbox{$^\prime$}}
\def \arcsec {\hbox{$^{\prime\prime}$}}

\author{
The VERITAS Collaboration:
E.~Aliu\altaffilmark{1},
S.~Archambault\altaffilmark{2},
T.~Aune\altaffilmark{3},
B.~Behera\altaffilmark{4},
M.~Beilicke\altaffilmark{5},
W.~Benbow\altaffilmark{6},
K.~Berger\altaffilmark{7},
R.~Bird\altaffilmark{8},
A.~Bouvier\altaffilmark{9},
J.~H.~Buckley\altaffilmark{5}
% ...
}

\altaffiltext{1}{Department of Physics and Astronomy, Barnard College, Columbia University, NY 10027, USA}
\altaffiltext{2}{Physics Department, McGill University, Montreal, QC H3A 2T8, Canada}
\altaffiltext{3}{Department of Physics and Astronomy, University of California, Los Angeles, CA 90095, USA}
\altaffiltext{4}{DESY, Platanenallee 6, 15738 Zeuthen, Germany; gernot.maier@desy.de}
\altaffiltext{5}{Department of Physics, Washington University, St. Louis, MO 63130, USA}
\altaffiltext{6}{Fred Lawrence Whipple Observatory, Harvard-Smithsonian Center for Astrophysics, Amado, AZ 85645, USA}
\altaffiltext{7}{Department of Physics and Astronomy and the Bartol Research Institute, University of Delaware, Newark, DE 19716, USA}
\altaffiltext{8}{School of Physics, University College Dublin, Belfield, Dublin 4, Ireland}
\altaffiltext{9}{Santa Cruz Institute for Particle Physics and Department of Physics, University of California, Santa Cruz, CA 95064, USA}
% ...
\begin{document}

\title{Long-term TeV and X-ray Observations of the Gamma-ray Binary HESS J0632+057}

%\title{Long-term observations at TeV energies of the variable Gamma-ray Source HESS J0632+057}
%POL: \title{Long-term TeV observations of the gamma-ray binary HESS J0632+057}

%\linenumbers

\begin{abstract} 
%-----------------------------------------------------------------------------------------------------------------------------------------------------

%
% POL: temptative draft:

HESS~J0632+057 is the only gamma-ray binary known so far whose position in the sky allows observations with
ground-based observatories both in the northern and southern hemispheres. Here we report on long-term observations of HESS~J0632+057
conducted with the VERITAS and H.E.S.S. Cherenkov Telescopes and the X-ray Satellite \emph{Swift}, spanning a time range from
2004 to 2012 and covering most of the system's orbit. The VHE emission is found to be variable, and is correlated with that at X-ray energies.
An orbital period of $315 ^{+6}_{-4}$ days is derived from the X-ray data set, which is compatible with previous  results, $P = (321
\pm 5$)~days. The VHE light curve shows a distinct maximum at orbital phases close to 0.3, or about 100 days after periastron passage,
which coincides with the periodic enhancement of the X-ray emission. Furthermore, the analysis of the TeV data shows for the first time a statistically
significant ($> 6.5 \sigma$) detection at orbital phases 0.6--0.9. The obtained gamma-ray and X-ray light curves and the correlation of the
source emission at these two energy bands are discussed in the context of the recent ephemeris obtained for the system. Our results are  compared
to those reported for other gamma-ray binaries.\\

% POL: lacking material:
%
% GM: I think this is ok, don't need to mention this.
%\noindent {\footnotesize{
%\bf{ {\it{NOT YET INCLUDED IN ABSTRACT:}
%positional agreement of VHE source with MWC 148; energy spectra (at different states); variability (orbit-to-orbit); 
% positional agreement of VHE source with MWC 148 --> really required in the Abstract?
% energy spectra (at different states); --> in X-rays as in Rea et al. ? --> to be done
% variability; --> orbit-to-orbit variability not possible at VHE. What about X-rays? --> to be done
%}}}}
\end{abstract}

\keywords{acceleration of particles â binaries: general - gamma rays: observations - individual (HESS J0632+057, VER J0633+057, MWC 148)}

\section{Introduction}

The very-high-energy (VHE; E$>$100 GeV) gamma-ray source HESS J0632+057 is a new member of the elusive class of gamma-ray binaries
\citep{Aharonian-2007, Hinton-2009, Bongiorno-2011}. These objects are characterized by a peak in their broad-band spectral energy distribution at
MeV-GeV energies, displaying variable high-energy emission as well as extended non-thermal radio structures. All known gamma-ray
binaries are high-mass X-ray binary systems, consisting of a compact object orbiting around a massive star of O or Be type. Besides
HESS J0632+057, only three binaries are clearly identified as VHE gamma-ray sources:  PSR~B1259-63/LS 2883 \citep{Aharonian-2005a}, LS
5039 \citep{Aharonian-2005b}, and LS I +61 303 \citep{Albert-2006,Acciari-2008}. In addition, some evidence for TeV emission has been
observed from Cygnus X-1 with the Major Atmospheric Gamma-ray Imaging Cherenkov telescope (MAGIC) \citep{Albert-2007}.
Finally, the High Energy Stereoscopic System (H.E.S.S.)
Collaboration recently reported the detection of a point-like source spatially coincident with the newly-discovered GeV gamma-ray
binary 1FGL J1018-5859 \citep{Fermi-2012,Abramowski-2012}, although no variability could be identified at TeV energies and the complex
morphology of the gamma-ray excess does not  yet allow an unequivocal association of the GeV and TeV sources.

%% (GM) Too detailed for the first paragraph of this paper
%The TeV data-set corresponded however to non-dedicated
%observations, and did not cover the whole source orbital phase range, preventing any clear sign of periodicity at VHE %energies.

%% (GM) I would like to remove this
%% Cyg X-1 is not a gamma-ray binary (emission not dominated in gamma rays and it was not a detection
% Finally, a hint of VHE flare from Cyg X-1 was reported by the Major Atmospheric Gamma-ray Imaging Cherenkov (MAGIC) Collaboration
% \cite{Albert-2007}, although further observations of this source at TeV energies could not yet confirm this signal.

Gamma-ray emission at VHE from HESS~J0632+057 was discovered serendipitously during observations of the Monoceros region in 2004-2005 
with H.E.S.S. \citep{Aharonian-2007}. Based on the point-like VHE gamma-ray appearance, X-ray variability, and spectral properties of the source,
\cite{Aharonian-2007} and \cite{Hinton-2009} suggested its identification as a new TeV binary system. HESS~J0632+057, located in the
direction of the edge of the star-forming region of the Rosette Nebula, was observed in the following years with the Very Energetic
Radiation Imaging Telescope Array System (VERITAS), H.E.S.S., and MAGIC telescopes. In 2006-2009, 
no significant emission was detected from the system
at energies
above 1 TeV with VERITAS  \citep{Acciari-2009}, suggesting its variability at VHE. In 2010 and 2011 clear gamma-ray
signals consistent with the initial H.E.S.S. results were observed \citep{Maier-2011,Aleksic-2012}, confirming the TeV variability. The
picture became clearer with the measurement of flux modulations with an initial period determination of ($321\pm5$)~days from multi-year observations in the
0.3--10 keV band with the {\it{Swift}} X-ray telescope (XRT) \citep{Bongiorno-2011, Falcone-2010}. The periodic modulation has been
recently confirmed by photometric measurements in the optical band \citep{Casares-2012}. The X-ray observations firmly established  the binary nature of
HESS J0632+057.

%HESS J0632+057 is located at  $6^{\mathrm{h}}32^{\mathrm{m}}58.3^{\mathrm{s}}, 
%+5^{\mathrm{o}}48'20''$ (R.A./Dec 2000) with 28'' statistical and 20'' systematic error on each axis. 

The optical counterpart of HESS J0632+057 is the massive B-star MWC 148  (HD~259440 = LS~VI~+05~11) at a distance of 1.1--1.7 kpc 
\citep{Aragona-2010}. MWC 148 is positionally coincident with the hard-spectrum and variable X-ray source  XMMU J063259.3+054801 \citep{Hinton-2009}.
At radio frequencies, weak emission was discovered at the position of  HESS J0632+057 with GMRT and VLA by \cite{Skilton-2009}. The radio
source is variable, but about 10 times fainter than the flux measured at similar  frequencies from other gamma-ray binaries (e.g.  LS 5039; \citealt{Moldon-2012}).
Observations with the European VLBI Network (EVN) show possible indications of an extended radio structure with a projected size of $\approx
75$ AU \citep{Moldon-2011}. Searches for pulsed emission from the system with {\it{Chandra}} and XMM-Newton in X-rays 
%and GBT in radio
yielded upper limits of  $\sim 22$--$48\%$ (depending on the frequency and emission state probed) on the pulsed fraction at energies 0.3--10~keV \citep{Rea-2011}. A potential association of HESS J0632+057
with the unidentified GeV gamma-ray source 3EG J0632+0521 has been suggested by \cite{Aharonian-2007}, although the source is marked
as ``possibly extended or multiple sources" and ``possibly source confused" in the third EGRET catalogue \citep{Hartman-1999}.
No gamma-ray emission has been reported by the \emph{Fermi} Large Area Telescope (LAT) Collaboration \citep{Caliandro-2012}
at MeV-GeV energies ($F_{100} < 3.0\times 10^{-8}$ ph~cm$^{-2}$~s$^{-1}$ 99\% C.L. upper limit above 100 MeV).

The spectral type of MWC 148 is B0pe  \citep{Morgan-1955}, characterized by an  optically-thick equatorial disk. The disk inclination is
uncertain with estimates ranging from $\geq 47^{\deg}$  to between $71^{\deg}$ and $90^{\deg}$
\citep{Casares-2012}.  Estimates of the physical parameters of the star have been reported, e.g., by \cite{Aragona-2010}  and
\cite{Casares-2012}, who derive an effective temperature of $T_{\mathrm{eff}}\approx 30000$ K, a mass of 13--19 $M_{\sun}$, and a radius of about 6--10
$R_{\sun}$. Orbital parameters of the binary system have been obtained through spectroscopic measurements  assuming an orbital period
of 321 days derived from the X-ray measurements \citep{Casares-2012}, including the orbit eccentricity $e = 0.83\pm0.08$,  phase of the
periastron $\Phi_0=0.967\pm0.008$ (defining phase 0 arbitrarily at $T_0 = \mathrm{MJD} \ 54857$), and an uncertain inclination of
$i\approx47^{\deg}$--$~90^{\deg}$. The large uncertainties involved  in this calculation lead to a broad range of masses allowed for the compact
object  ($M_c\approx 1.3-7.1 \ M_{\sun}$). Its nature, neutron star or black hole, is therefore unclear.

The physical processes leading to particle acceleration and gamma-ray emission in binaries are under debate. Two major classes of models are usually invoked to
explain their high-energy emission (see, e.g., \citealt{Mirabel-2012}). In the first one, acceleration of charged particles takes place in accretion-powered
relativistic jets (so called {\it microquasars}; \citealt{Taylor-1984, Mirabel-1994}) and usually implies a black hole as compact object.
In the second one, high-energy emission is produced by the
ultra-relativistic wind of a rotation-powered energetic pulsar, either scattering directly off the photon field of the companion star or photons from a  circumstellar disk 
\citep{Ball-2000, vanSoelen-2011,Khangulyan-2012}, or accelerating particles in the region where the pulsar wind collides with the disk material or the wind of the massive companion \citep{Maraschi-1981, Dubus-2006a}. Note that the shocked wind material could also be relativistic in this last case \citep{Bogovalov-2008,Dubus-2010}.

The interactions between the compact object, the massive star, and their winds and magnetic fields form a complicated environment, in which 
acceleration, radiation, and absorption processes take place (see, e.g., \citealt{Bednarek-2011}). This complex nature of gamma-ray binaries may lead to the
variety of emission patterns observed in these systems. In particular, two of the best studied gamma-ray binaries show emission that is
modulated by their orbital period  (PSR~B1259-63, \citealp{Aharonian-2005a}; LS 5039, \citealp{Aharonian-2005b}), while in the case of the system
LS~I~+61~303 the source displays both periodic and episodic variability (\citealp{Albert-2009,Acciari-2011}; see also
\citealp{Chernyakova-2012}).

Below, long-term gamma-ray observations of HESS J0632+057 with the VERITAS and  H.E.S.S. facilities are reported, and X-ray observations with
the {\it{Swift}}-XRT telescope.  Section \ref{Observations} describes the VHE and the X-ray data-sets, including VHE observations taken up to
February 2012 and {\it{Swift}}-XRT observations up to March 2012. These results provide for the first time a wide coverage of the system's
orbital phases. Section \ref{Results} is focused on the results obtained from the VHE and X-ray data analysis. 
These results are discussed in Section \ref{Discussion} in a multi-wavelength context, and compared to those obtained from other well-studied 
gamma-ray binaries.

\section{Observations}
\label{Observations}
%-----------------------------------------------------------------------------------------------------------------------------------------------------

VERITAS and H.E.S.S. are ground-based imaging atmospheric Cherenkov telescopes (IACTs) built to detect the faint flashes of Cherenkov light from air
showers initiated in the atmosphere by high-energy gamma-ray photons. The instruments are very similar in their performance with effective areas of
$\gtrsim 10^5$ m$^2$ over an energy range from $\sim $100 GeV to 30 TeV, energy resolution $\sim 15-20$\%, and angular resolution
$\approx0.1^{\deg}$. The high sensitivities of H.E.S.S. and VERITAS enable the detection of sources with a flux of 1\% of the Crab Nebula in less
than 30 hours of observations.

The analysis of the VHE data from the two instruments follows similar initial steps, consisting of calibration, image cleaning, and second-moment parametrization
of Cherenkov shower images \citep{Hillas-1985}, which provide the reconstruction of the shower direction, energy, and impact parameter using stereoscopic methods
(see, e.g.,~\citealt{Krawczynski-2006}). For the H.E.S.S. data analysis, a further fitting procedure is employed, for which the Hillas parameters are used as the
starting point for a refined derivation of the shower parameters based on a log-likelihood comparison of the raw, uncleaned image with a pre-calculated shower
model \citep{Naurois-2009}.  A shower event must be imaged by at least two out of four telescopes to be used in the VHE analysis of data from both
instruments,  and additional cuts on the shape of the event images and the direction of the primary particles are used to reject the far more numerous background
events. Most of the VHE data were taken in wobble mode in both the VERITAS and H.E.S.S. data-sets, wherein the source is positioned at an offset from the camera
centre of about 0.5$^{\deg}$ to allow for simultaneous and symmetric background regions to be used during the data-analysis procedure. All results
presented here have been cross-checked with independent analysis chains.

\subsection{VERITAS VHE Gamma-Ray Observations}
%-----------------------------------------------------

The VERITAS observatory is an array of IACTs located at the Fred Lawrence Whipple Observatory in southern Arizona (1300~m above sea level, N$31\deg
40\arcmin 30\arcsec$, W$110\deg 57\arcmin 08\arcsec$).  The mirror area of each telescope is 110 m$^2$ and the total field of view (FoV) of
the instrument is 3.5$^{\deg}$ in diameter.

% note: pre-summer 2007 data is not in loggen!

VERITAS observed the sky around HESS J0632+057 for a total of 162 hours between December 2006 and January 2012 (see Table
\ref{table:ObservationsAndResults} for details). A total of 144 hours of observations passed quality-selection criteria, which remove data taken
during bad weather or with technical problems. The instrument went through several important changes during this period. The data from December
2006 were taken during the construction phase of VERITAS with three telescopes only.  The array was completed in September 2007  with four
telescopes in total.  In September 2009 the array layout was improved by moving one telescope, leading to an improved sensitivity, which makes
it possible to detect point-like sources with a flux of 1\% of the Crab Nebula in less than 30 hours of observations (1 h for 5\% of the Crab
Nebula flux). The corresponding values prior to 2009 are less than 50 hours and 2 h for sources with  1\% and 5\% of the flux of the
Crab Nebula respectively.

Observations with VERITAS are possible during dark sky and moderate moonlight conditions (moon illumination $<35$\%). The elevated background light
levels during moderate moonlight lead to a lower sensitivity to gamma rays near the low energy threshold of the instrument. Observations were
performed in a zenith angle range of 26$^{\deg}$--40$^{\deg}$. All  VERITAS observations, with the exception of the observations in December 2006 -
January 2007, were taken at a fixed offset of 0.5$^{\deg}$ in one of four directions. The energy threshold\footnote{The energy threshold is defined
as the position of the peak of the differential energy spectrum (assuming a power law for the spectral shape; here $\Gamma=-2.5$ was assumed) of the source
convolved with the effective area curve of the detector.} after analysis cuts, where a cut on the integrated charge per image of 500 digital counts 
($\approx 90$ photoelectrons) is applied, is 230 GeV for the data set presented here (average elevation angle of
62$^{\deg}$). For more details on the VERITAS instrument see, e.g., \cite{Acciari-2008}. The extraction region for photons from the putative
gamma-ray source is defined by a 0.09$^{\deg}$ radius circle centered on the position of the X-ray source  XMMU J063259.3+054801 (coincident with the
star MWC 148; \citealp{Hog-1998}). The background in the source region is estimated from the same FoV using the ring-background model with a ring
size of 0.5$^{\deg}$ (mean radius) and a ring width of 0.1$^{\deg}$ \citep{Berge-2007}. In order to reduce systematics in the background estimation,
regions around stars with B-band magnitudes brighter than 6 are excluded from the background control regions.

\subsection{H.E.S.S. VHE Gamma-ray Observations}
%----------------------------------------------------

The H.E.S.S. observatory is located in the Khomas highland of Namibia (1800~m above sea level, S$23\deg 16\arcmin 18\arcsec$,
E$16\deg 30\arcmin 00\arcsec$). The H.E.S.S. array consists of four 13\,m diameter telescopes positioned in a
square of side length 120\,m. Each telescope is equipped with a tessellated spherical mirror of 107~m$^{2}$, focusing the Cherenkov
flashes onto a camera that covers a FoV of about 5$^{\deg}$ in diameter. For a detailed description
of the system, see \cite{Aharonian-2006} and references therein. No data from the H.E.S.S.-II array, which includes the addition of a
central, 28\,m diameter telescope, are included in this publication.

HESS\,J0632+057 was observed yearly with H.E.S.S. from 2004 until 2012. The FoV around the source was initially covered by deep observations of the
Monoceros Loop SNR/Rosette Nebula region, in the search for potential sources of VHE emission, including also two unidentified EGRET sources. After
the discovery of HESS\,J0632+057 \citep{Aharonian-2007}, further dedicated observations were obtained to better constrain its nature, in particular
searching for TeV variability/periodicity following the non-detection of the source reported by the VERITAS Collaboration in 2006 to 2009
\citep{Acciari-2009}. The total acceptance-corrected effective exposure time on HESS\,J0632+057, including both the initial discovery dataset and the
following pointed observations, consists of 53.5 hours of data after standard quality selection cuts \citep{Aharonian-2006}. Observations were
performed over a large range of elevation angles (32$^{\deg}$--62$^{\deg}$, with an average of 56.0$^{\deg}$). The data have been analyzed using the
Model Analysis technique \citep{Naurois-2009} and  cross checked with a Hillas-based analysis, making use of an independent calibration procedure of
the raw data, providing compatible results. The results presented here, using standard cuts where a cut of 60 photo-electrons on the intensity of the
extensive air showers is applied, provide a mean energy threshold of $\sim 220$~GeV for the dataset presented.

%Cross-checks have been performed using a a multivariate analysis
%\citep{Becherini-2011} and a Model Analysis \citep{Naurois-2009}, including independent calibration of pixel
%amplitudes and identification of problematic or dead pixels in the IACTs cameras, leading to compatible results.

\subsection{\emph{Swift} X-ray observations}
%--------------------------------------------------------------------------------------------------------------------------------------------

The \emph{Swift} X-ray Telescope (XRT), which is sensitive in the 0.3 -- 10 keV band \citep{burrows05}, was used to monitor HESS J0632+057 during
the time period from 26th January 2009 to 15th February 2012. The observations had typical durations of $\sim4-5$ ksec. The temporal spacing
between observations is typically $\sim$1 week, although some time periods include several month-long time gaps due to observing constraints,
whilst others include daily observations. 

The \emph{Swift}-XRT data were processed using the most recent versions of the standard \emph{Swift} tools: \emph{Swift} Software version 3.9
and HEASoft FTOOLS version 6.12 \citep{blackburn95}. Observations were processed individually using {\it{xrtpipeline}} version 0.12.6. Hot and
flickering pixels were  removed using {\it{xrthotpix}}, and hot pixels were additionally removed by rejecting data where the XRT CCD
temperature is $\geq -47\deg$~C. Standard grade selections of 0--12 were used for these Photon-Counting (PC) mode data.

Light curves were generated using {\it{xrtgrblc}} version 1.6. 
Circles were used to describe the source regions. The source count rate was always $<$ 0.1 counts s$^{-1}$, so pile-up correction was not needed.
Annuli with inner radii outside the radii of the source regions were used to describe the background
regions for all data. The radii of the regions depend on the count rate in each temporal bin. Point-spread-function corrections and corrections for the relative sizes of the
extraction regions were applied. For light-curve plotting, each observation results in one bin.

\begin{figure}
\plotone{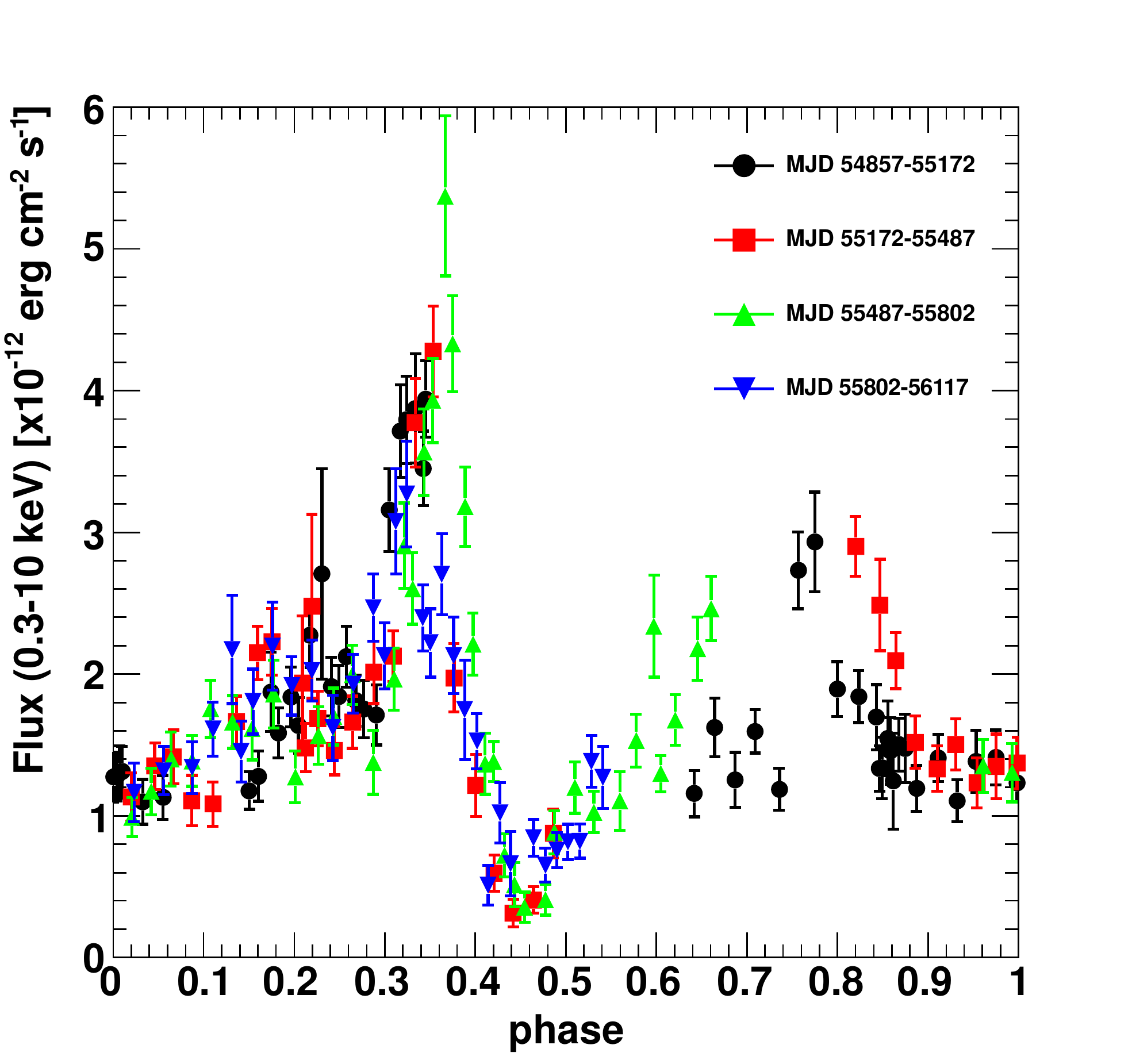}
\caption{\label{fig:XRT-315d}
Phase-folded X-ray (0.3--10 keV) light curve assuming an orbital period of 315 days. 
Different orbital cycles are indicated by  markers and colors.
Vertical error bars show $1\sigma$ statistical uncertainties.
}
\end{figure}

In order to calculate count-rate to flux conversion factors, spectral fits were generated using XSPEC version 12.7.1 \citep{arnaud96}. Since the
spectral shape may vary, we defined high, medium and low states based on the rate light curve, and binned data together within each subset to perform
a spectral fit. The high state contains only data from the large peak (see Fig.~\ref{fig:XRT-315d}), approximately at phases 0.32--0.39. The
low state contains only data during the ``dip" that immediately follows each large peak, roughly from phase 0.42--0.48. The medium state data
contains data from phase 0.6 to phase 0.32 (see Table \ref{table:SwiftSpectralResults}). 
%An absorbed power-law was used to fit
%the spectra from each of these three data subsets, using $\chi^2$ statistics and fixing the absorption parameter,
%$N_{\rm H}$, to 3.81$\times10^{21} $ cm$^{-2}$, which is the value found from fitting the entire data set with $N_H$ left as a free parameter. 
An absorbed power-law was used to fit the spectra from each of these three data subsets, using $\chi^2$~statistics and the {\it{XSpec}} photoelectric model {\it{wabs}} to define absorption cross sections and abundances  \citep{Mor83}. 
For the combined data set, $N_H$ was left as a free parameter and converged to $3.81 (+0.29,-0.27) \times 10^{21}$ cm$^{-2}$. For each of the data subsets, we fixed the absorption parameter, $N_H$, to $3.81 \times 10^{21} $cm$^{-2}$.
After
performing a spectral fit of an absorbed power-law to each data subset, the unabsorbed flux was calculated to create three separate rate-to-flux
conversion factors.

%Fig. \ref{fig:XRT-MJD} shows the long-term X-ray data set of XMMU J063259.3+054801, from January 26th 2009 to February 12th 2012 , as
%measured by \emph{Swift}-XRT. Fig.~\ref{fig:SED} shows the X-ray spectra in each of the three states described above.

%% POL: should´nt this last paragraph be in the Results section?

\section{Results}
%--------------------------------------------------------------------------------------------------------------------------------------------
\label{Results}

\begin{figure}
\plotone{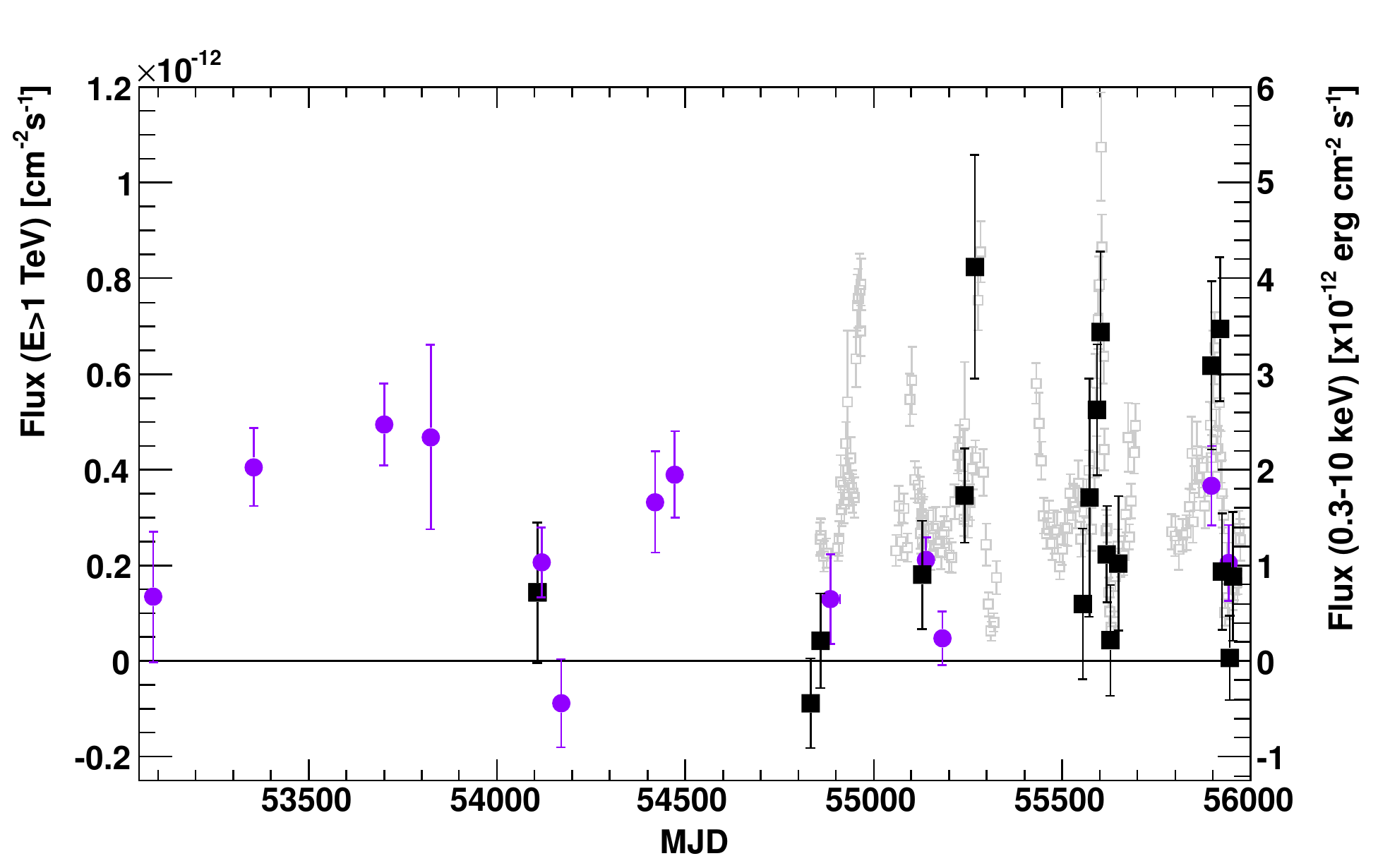}
\caption{\label{fig:XRT-MJD}
Long-term observations of HESS J0632+057 with H.E.S.S. (round purple markers) and VERITAS (square black markers) at energies $>1$ TeV
and in X-rays with \emph{Swift}-XRT (0.3--10 keV; open grey markers)  from March 2004 to February 2012.
}
\end{figure}

\begin{figure}
\plottwo{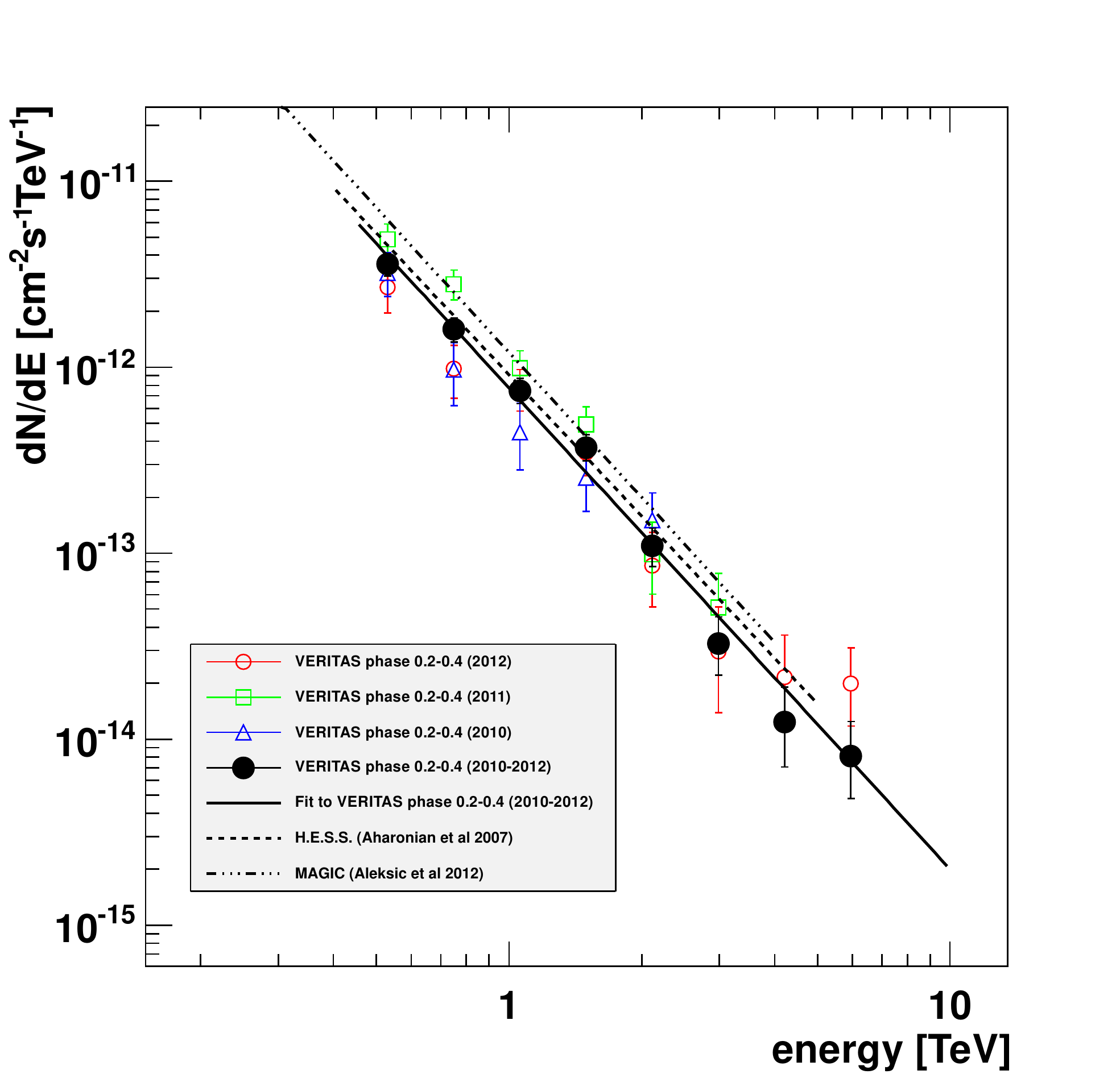}{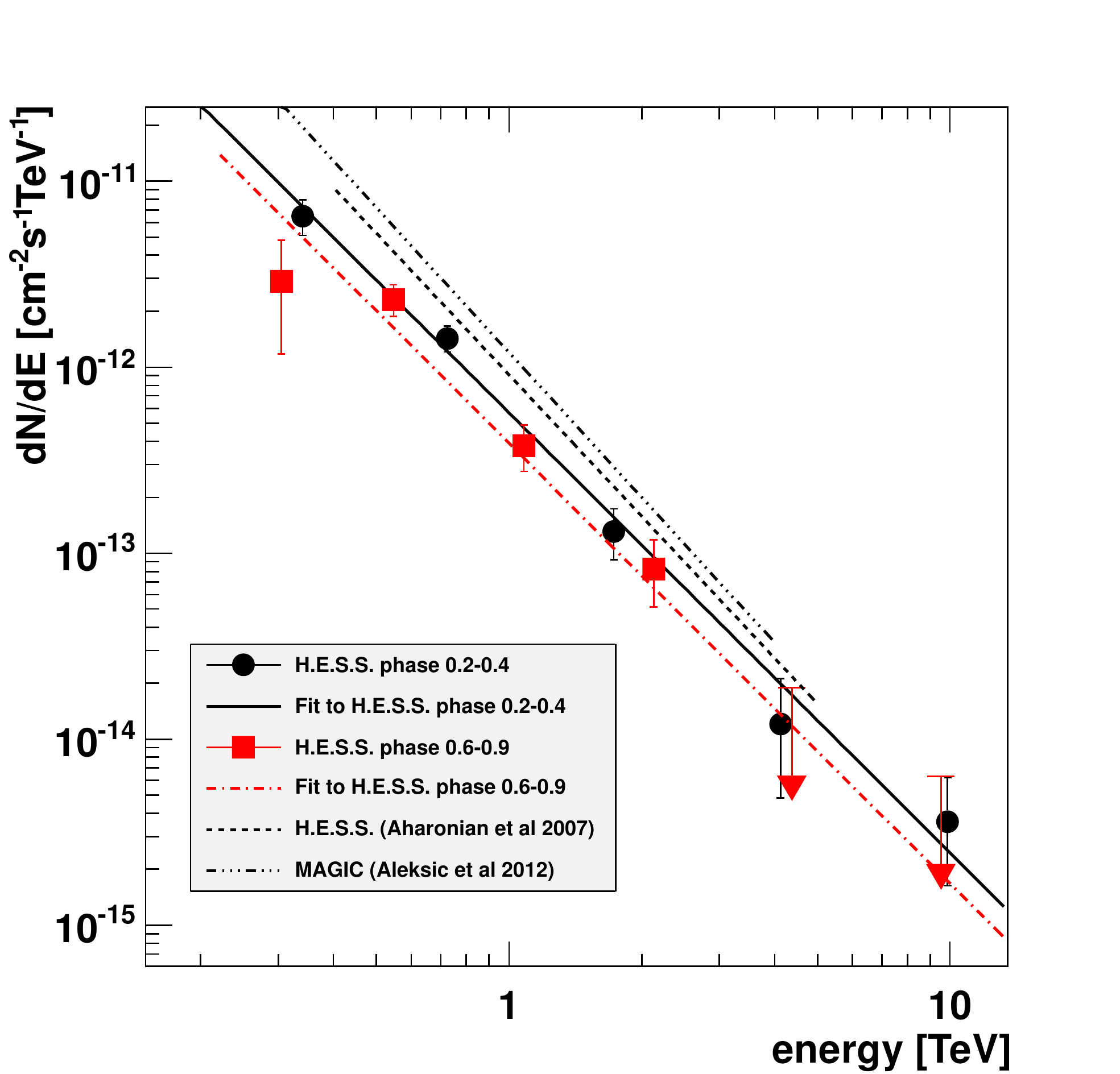}
\caption{\label{fig:VTS-ESpec}
Differential energy spectra of VHE photons for HESS J0632+057 as measured by VERITAS and H.E.S.S. The continuous lines show the results from fits
assuming a power-law distribution of the data and a spectral index of -2.5.  Fit results can be found in Table \ref{table:EnergySpectrum}. Vertical
error bars indicate $1 \sigma$  statistical errors.  Left:  Differential energy spectrum for orbital phases 0.2 to 0.4 as measured by VERITAS during
different orbits. Right:  Average differential energy spectrum for orbital phases 0.2 to 0.4 and 0.6 to 0.9 as measured by H.E.S.S. For comparison,
although taken at different orbital phases, the energy distributions as published by the H.E.S.S.  (\citealt{Aharonian-2007}; measurement in December
2004 and December 2005) and MAGIC Collaborations (\citealt{Aleksic-2012}; measurement in February 2011) are indicated by the dashed and
dotted-dashed lines, respectively.
}
\end{figure}

HESS~J0632+057 was detected as a source of VHE gamma rays at a high confidence level by both observatories.  The detection significance  of the
highly variable gamma-ray source derived from 144 hours of VERITAS observations is 15.5 $\sigma$, whereas the source is detected with H.E.S.S. at a
significance of 13.6 $\sigma$ in a total of 53.5 hours of observations. Figure~\ref{fig:XRT-MJD} summarizes these results by showing the long-term
gamma-ray light curve for energies above 1 TeV for all H.E.S.S. and VERITAS measurements from 2004 to 2012 (detailed results of these observations
can be found in Table \ref{table:ObservationsAndResults}). All significances, fluxes, and spectral analyses are calculated using the X-ray source XMMU
J063259.3+054801  position \citep{Hinton-2009}. Integral fluxes above 1~TeV are calculated assuming a spectral distribution following a power law with
a photon index of -2.5 (see Fig.~\ref{fig:VTS-ESpec} for the measured differential energy spectrum of HESS J0632+057).

% calculation of variability index as in 1st Fermi LAT catalogue
% assuming 20% systematic uncertainty in Flux
% Chi2 = 94.773 with 29 deg of freedom, prob is 6.6e-9
% (output from plotLightCurvesMJD_VHE_Xrays( false )
% Chi2=122.6221 with 30 deg of freedom, prob is 3.6990e-13

The source has been found to be variable, as earlier measurements with H.E.S.S. and VERITAS suggested \citep{Acciari-2009}. A calculation of the
variability index $V$, a  $\chi^2$-criterion described in detail in \citet{Abdo-2010}, results in $V=94.7$ for the combined H.E.S.S./VERITAS light
curve with 30 flux points and the assumption of a systematic error on the flux estimation of each instrument of 20\% \citep{Aharonian-2006}. This
means that  the light curve is significantly different from a constant one at a confidence level of $1-3.6\times 10^{-13}$  (7.1 $\sigma$). A search
for variability patterns and an orbital period in the VHE data is unfortunately hampered by  insufficient coverage and large gaps in the light
curve.  A much larger data set with denser sampling of the orbital period is needed to derive this from observations of HESS J0632+057. 

The high-significance detection allows  the position of the VHE source to be updated using both the VERITAS (from 2010 to 2012) and H.E.S.S. (from
2004 to 2012) data sets.  The best fit position from the VERITAS subset is
RA=$06^{\mathrm{h}}33^{\mathrm{m}}0.8^{\mathrm{s}}\pm0.5^{\mathrm{s}}_{\mathrm{stat}}$ and DEC=$+5^{\deg}47\arcmin
39\arcsec\pm10\arcsec_{\mathrm{stat}}$  (J2000 coordinates) with a source extension of the gamma-ray image compatible with the VERITAS gamma-ray
point spread function\footnote{The object is added to the  VERITAS source catalogue with the name VER J0633+057.}. The systematic uncertainty in the
position due to telescope pointing errors is below $50\arcsec$. This subset of VERITAS data, with observation dates after the optimization of the
array layout, has been chosen to minimize systematic errors on the direction determination. The best-fit position for HESS~J0632+057 resulting from
the new H.E.S.S. measurements is RA=$06^{\mathrm{h}}32^{\mathrm{m}}59.4^{\mathrm{s}}\pm1.1^{\mathrm{s}}_{\mathrm{stat}}$ and DEC=$+5^{\deg} 47\arcmin
20\arcsec\pm16.1\arcsec_{\mathrm{stat}}$  (J2000). The positional agreement between the updated VERITAS and H.E.S.S. position, the original H.E.S.S.
detection \citep{Aharonian-2007}  and the X-ray source XMMU J063259.3+054801, as well as its compatibility with a point-like source, have thus been
confirmed.

The long-term X-ray light curve of XMMU J063259.3+054801, from 26th January 2009 to 12th February  2012, as measured with \emph{Swift}-XRT is shown
together with the VERITAS and H.E.S.S. measurements in Fig.~\ref{fig:XRT-MJD}.  The X-ray light curve is highly variable with several distinguishable
features appearing periodically. The analysis reported here follows closely that in \citet{Bongiorno-2011}, but using one additional year of data.
Z-transformed discrete correlation functions are applied to determine the overall variability patterns in the X-ray light curve and the correlation
between X-ray and gamma-ray emission (the number of data points in the gamma-ray light curve is not sufficient for an autocorrelation analysis).  The
Z-transformed discrete correlation functions (Z-DCF) \citep{Alexander-1997} are based on the discrete correlation analysis developed by
\cite{Edelson-1988}, employing  additionally equal population binning and Fisher's z-transform, that transforms the correlation coefficient into an
approximately normally distributed variable.  This leads to a more robust estimation of the correlation coefficients. Errors on the Z-DCF
coefficients are calculated in this analysis with a Monte Carlo-based approach using 10.000 simulated light curves with flux values randomly changed according
to  their measurement uncertainties and assuming them to be normally distributed. Time lags and their 68\% fiducial intervals are calculated  from
the peak likelihood of the Z-DCF using Bayesian statistics. Z-DCF have been used in preference to Pearson's correlation coefficient as the latter
does not take uncertainties on the flux values into account.

\begin{figure}
\plotone{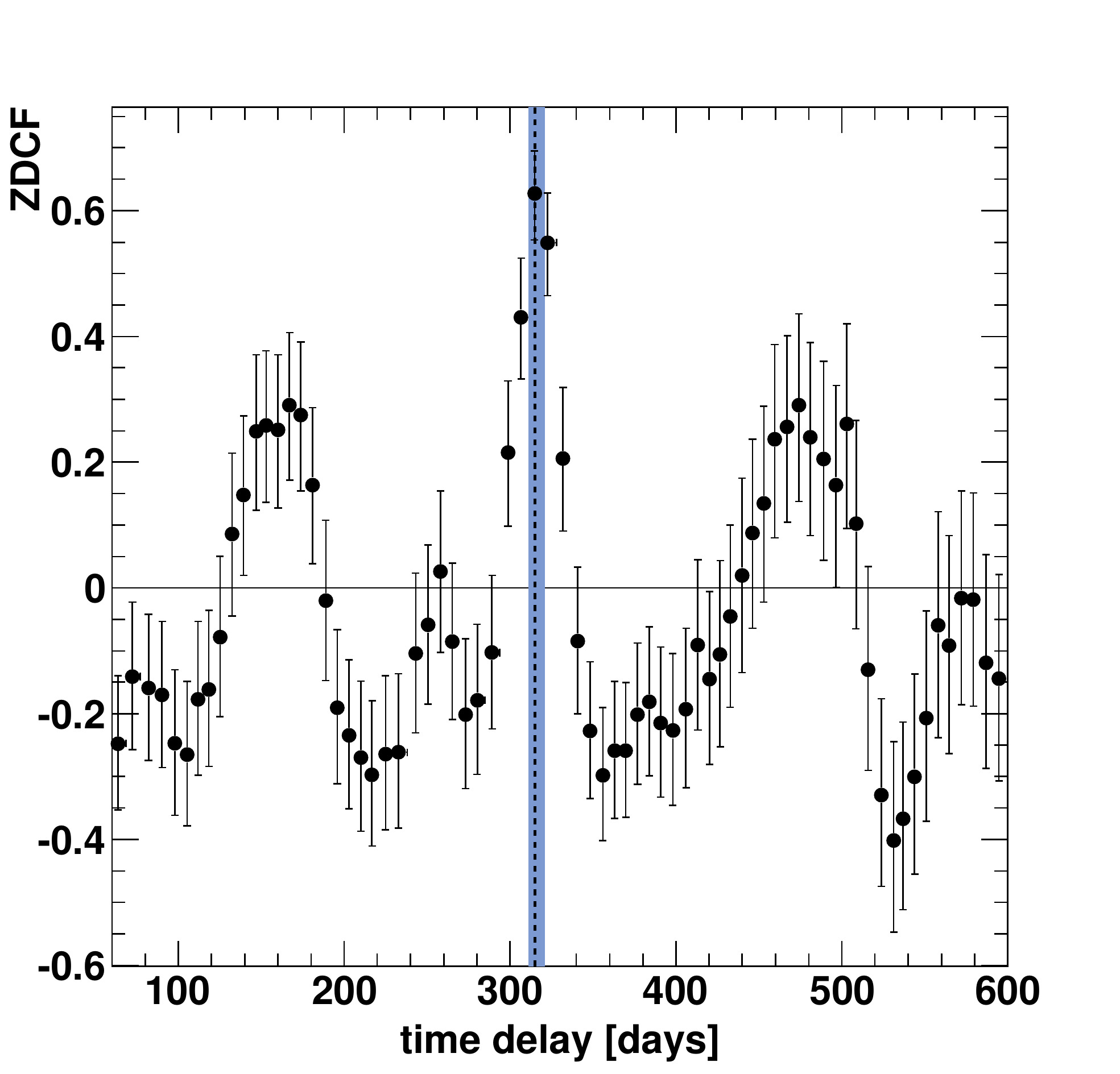}
\caption{\label{fig:XRT-ZDCF}
Z-transformed discrete autocorrelation function (Z-DCF) for the \emph{Swift}-XRT light curve shown in Fig. \ref{fig:XRT-MJD}.  The errors
bars denote the $1\sigma$ sampling errors resulting from a MC-based error calculation as described in the text. The dashed line and the
blue band indicate the most likely modulation period of  $315^{+6}_{-4}$ days and the corresponding 68\% fiducial interval.}
\end{figure}

Figure~\ref{fig:XRT-ZDCF} shows the results from the autocorrelation analysis of the X-ray light curve.  Flux modulation with a period of ($321\pm5$)
days has been presented in \citet{Bongiorno-2011}, applying peak-fitting algorithms and Z-DCFs to a subset of the \emph{Swift}-XRT data presented in
this paper. The larger data set available now (154 compared to 112 flux points used in \citealp{Bongiorno-2011}) results in a compatible period
of $315^{+6}_{-4}$ days. We use therefore in this paper the following phase definition: MJD$_0=54857$ (arbitrarily set to the date of the first
\emph{Swift} observations) and  period $P=315$ days. It should be noted that the orbital parameters as derived from \citet{Casares-2012} remain
approximately unchanged  by this small change in orbital period (\cite{Casares-2012a}, private communication).

The phase-folded X-ray light curves as shown in Fig.~\ref{fig:XRT-315d} illustrate the very regular
emission pattern of HESS~J0632+057  with a strong maximum at phases $\sim$0.35, a marked dip at phases
$\sim$0.45 and an intermediate flux level at orbital phases $\sim$~0.6--0.3. There are also
indications of a second maximum at phases $\sim$~0.6--0.9, with a flux level about half of
that of the main peak at phases $\sim$0.35.  Apart from this very regular pattern, orbit-to-orbit
variability at X-ray energies is also visible, e.g., around the region of the emission maxima. 

% phase-folded gamma-ray light-curve

\begin{figure}
\plotone{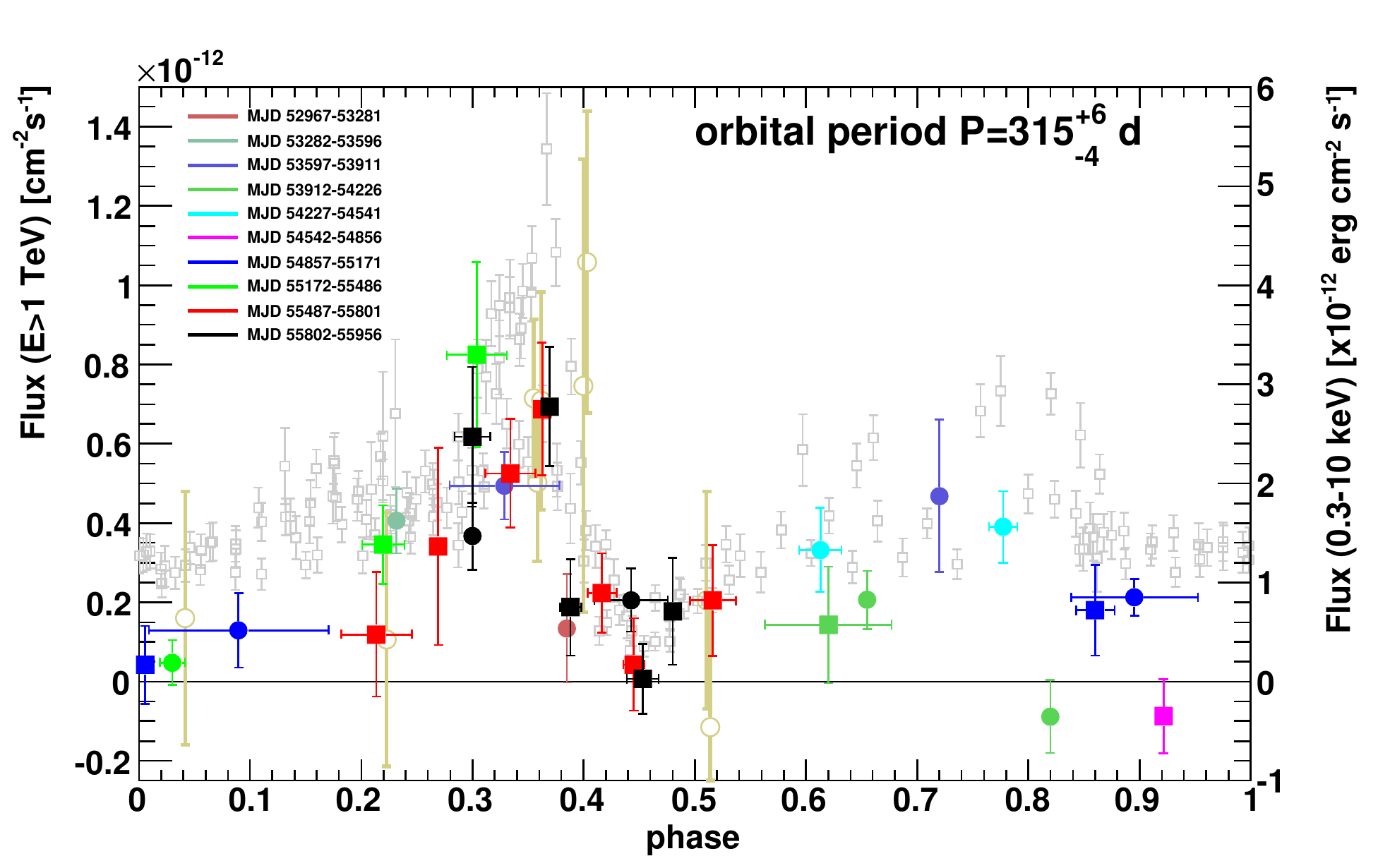}
\caption{\label{fig:VHE-XRT}
Integral $\gamma$-ray fluxes above 1 TeV (vertical scale on the left) from H.E.S.S. (filled round markers), MAGIC (brown round open markers, 
scaled to 1 TeV assuming a power law with index $-2.6$; \citealt{Aleksic-2012}) 
and VERITAS (filled squared markers).
X-ray fluxes (0.3--10 keV) are shown as measured by \emph{Swift}-XRT  (open square grey marker; vertical scale on the right).
All measurements are folded with the orbital period of 315 days; the colors indicate different orbits.
Vertical error bars show $1\sigma$ statistical uncertainties, horizontal error bars indicate the width of the corresponding observing interval.
}
\end{figure}

\begin{figure}
\plottwo{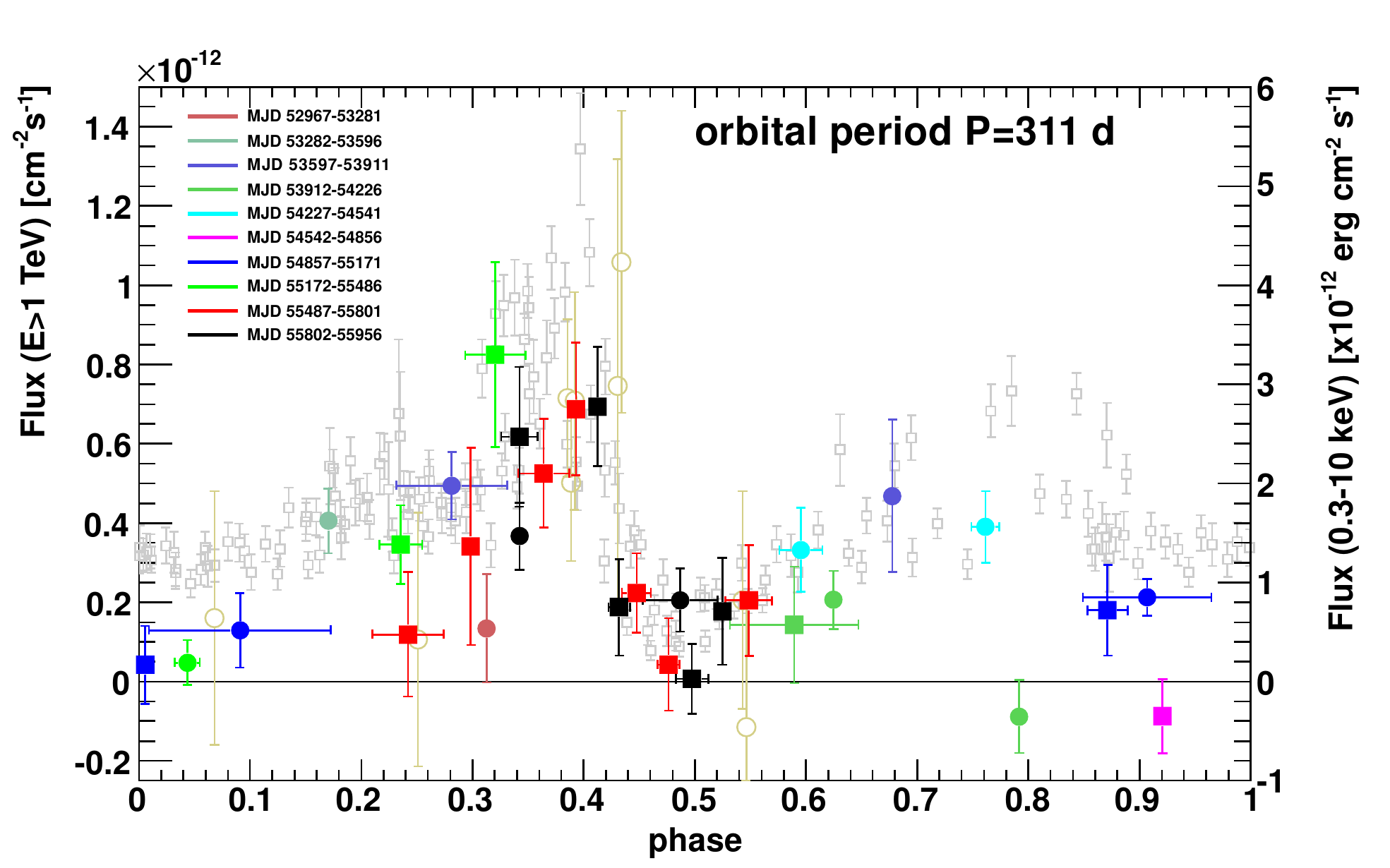}{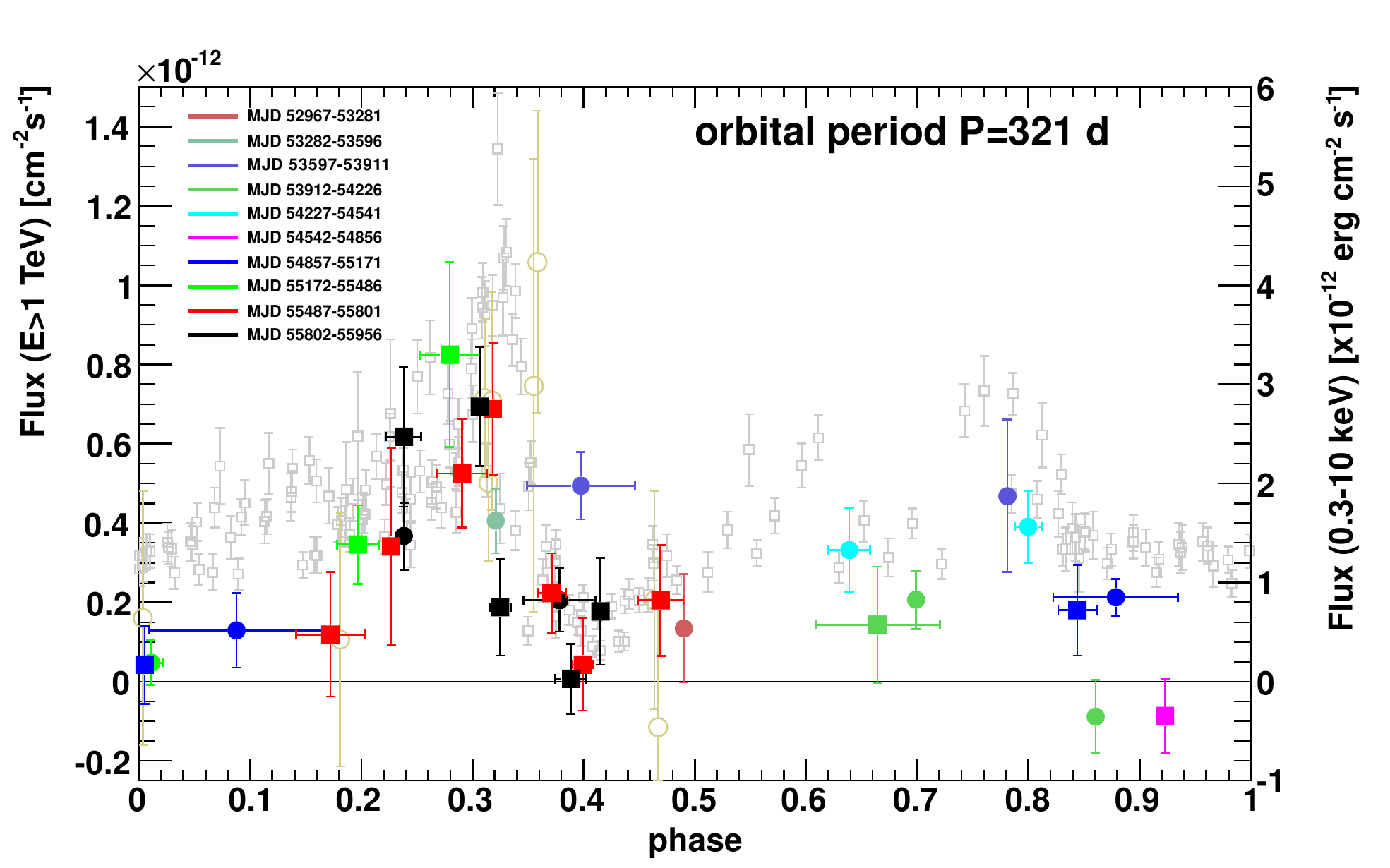}
\caption{\label{fig:VHE-XRT-1sigma}
Impact of the $1\sigma $ error of the orbital phase on the phase-folded light curves. %(orbital period is determined to be $p=315^{+6}_{-4}$ days). 
Phase-folded light curve using an orbital period of 311 (left) and  321 (right) days. The maximum of the X-ray light-curve coincides with that at TeV energies taking both values for the period and occurs at orbital phases $\sim 0.3$. Enhanced X-ray emission at phases $\sim 0.6$--0.9 is also observed in the two panels, for which most of the corresponding TeV data points leading to the source detection in this phase interval are still included.
See Fig. \ref{fig:VHE-XRT} for a description of the figure axes, markers and error bars.
}
\end{figure}

The gamma-ray light curve has been folded with the  orbital period derived from the X-ray data. The uncertainty in the orbital period translates into
a noticeable uncertainty in orbital phase, since the VHE observations presented here are taken over a period of $\sim$8 years. In order to ensure
that none of the conclusions presented in the following depends on the particular value  of the orbital period, we present in Fig.~\ref{fig:VHE-XRT}
the gamma-ray light curve folded by a period of 315 days, while in Fig.~\ref{fig:VHE-XRT-1sigma} periods of 321 and 311 days have been
applied. The phase-folded light curves reveal several important characteristics of the high-energy emission: a clear detection of the source in the
phase range 0.2--0.4, around the maximum of the X-ray light curve, with a flux of 2-3\% of that of the Crab Nebula;  a first detection of a gamma-ray
emission component at orbital phases in the range 0.6--0.9, in which a secondary peak in the X-ray light curve is also observed; and a non-detected
low state at all other orbital phases.

It should be noted that the H.E.S.S. dataset leading to the detection of VHE emission in orbital phases 0.6--0.9 comprises observations taken at different
epochs, from March 2006 to October 2009 (see Table \ref{table:ObservationsAndResults} for details). HESS~J0632+057 is detected at a significance level of
7.7$\sigma$ when all data falling in this phase interval are considered.  
%
%A hint of emission at this phase interval from the analysis of the H.E.S.S data was
%already reported by \citep{Bordas-2012}. The inclusion of new data not included
%before, for which an updated callibration was required, is reported here for the
%first time and is responsible for the improved detection significance of the source
%in this phase-range. We remark that the significances reported here are pre-trials.
%It has to be noted however that a search for TeV emission has been performed in the
%phase range 0.6--0.9 as defined by the presence of a second/smaller X-ray
%high-state, so this {\it{prior}} information prevents from the need for a trials
%factor penalty to the significance estimate. 
%
%
The search for TeV emission has been performed in this phase range as defined by the presence of a second/smaller X-ray high-state.
Therefore, no trial factor penalty needs to be applied to the significance estimation.\footnote{ \footnotesize{An upper limit on the
number of trials for a blind search of significant emission in the phase-folded light curve can nevertheless be derived as follows: The
light curve is first divided into 10 bins, and a significant detection is then evaluated for all intervals of  0.1, 0.2, 0.3
...0.9 and 1.0 width in orbital-phase, without repetition (e.g. intervals of 0.3 width, one has to consider the cases 0.0--0.1 + 0.1--0.2 +
0.2--0.3, 0.0--0.1 + 0.1--0.2 + 0.3--0.4 ... 0.7--0.8 + 0.8--0.9 + 0.9--1.0.) In each case, a number of trials $= 10!/[n!(10-n)!]$ is
obtained, where $n = 1, 2, 3...9$. The total number of trials resulting from this computation is $1275$. A 7.7$\sigma$ detection would
be reduced to $\sim 6.7\sigma$ in this extreme case using $P_{\rm t} = 1.0 - (1.0 - P)^{N_{\rm trials}}$ ($P_{\rm t}$, and $P$ are the
pre- and post-trial probabilities, respectively, and $N_{\rm trials}$ is the number of trials), which is still highly significant. Note 
that this is a true lower limit on the detection significance, as we do not require continuity when merging different phase intervals
 which would further reduce the total number of trials.}} 
 Moreover, the variability index for the 18 VHE flux points outside of the phase bins around the main maximum (phases 0.2--0.4) has been
computed. This calculation yields a value for the variability index of $V=52.3$, corresponding to a probability of $2\times10^{-5}$,
suggesting that there may exist variations in the source VHE light curve away from the main peak. A likelihood ratio test was also
performed to further explore if the detection in phases 0.6--0.9 constitutes a significant secondary maximum above the baseline level.
This baseline flux is computed by adding all data in the phase ranges 0.0--0.2, 0.4--0.6, and 0.9--1.0. The likelihood function is then
defined as a product of two Gaussian distributions of flux measurements $\phi_{0.6-0.9}$ and $\phi_{\rm base}$ for the emission in
phases 0.6--0.9 and in the baseline range, respectively, stating that $\phi_{0.6-0.9}$ is a factor $K_{0.6-0.9}$ times higher than 
$\phi_{\rm base}$. The likelihood-ratio test provides a value for $K_{0.6-0.9}$ in the range $[0.83, 3.90]$ at a 99.7\% (3$\sigma$)
confidence level, with a best fit value $K_{0.6-0.9} = 1.72$. Therefore, although the emission at orbital phases 0.6--0.9 is higher
than the baseline flux, it cannot be claimed as a secondary peak with the present dataset at a high confidence level.

%
%Therefore, a proper characteirzation of the emission at orbital phases 0.6--0.9 as a secondary peak in the lightcurve above this significance can not be %ensured within our data-set, for which further VHE observations may be required.
%Moreover, to further characterize the detected VHE emission in this phase range and ensure that it constitutes a second maximum in the light curve, and not part of a constant baseline VHE flux, we computed the variability index for the 18 VHE flux points outside of the phase bins around the main maximum (phases 0.2--0.4). This calculation yields a value for the variability index of $V=52.3$, corresponding to a probability of $2\times10^{-5}$, pointing therefore towards a true double peak structure of the TeV light curve, with a strong maximum at phases 0.2--0.4 and a weaker one at phases 0.6--0.9.
%
%This detection is still robust at the level of xx sigma if one takes an overly conservative estimate of the significance by assuming a trials factor defined by the number of 0.6-0.9 phase bins available for a random search for emission (i.e. trials factor = 1/(0.9-0.6) = 3.33)
%
%A detection of TeV emission, at the 7.7 sigma significance level, was found in a search for TeV emission in the phase range defined by the second/smaller X-ray high state, i.e. phase 0.6-0.9.  This detection is still robust at the level of xx sigma if one takes an overly conservative estimate of the significance by assuming a trials factor defined by the number of 0.6-0.9 phase bins available for a random search for emission (i.e. trials factor = 1/(0.9-0.6) = 3.33)
%

\begin{figure}
\plottwo{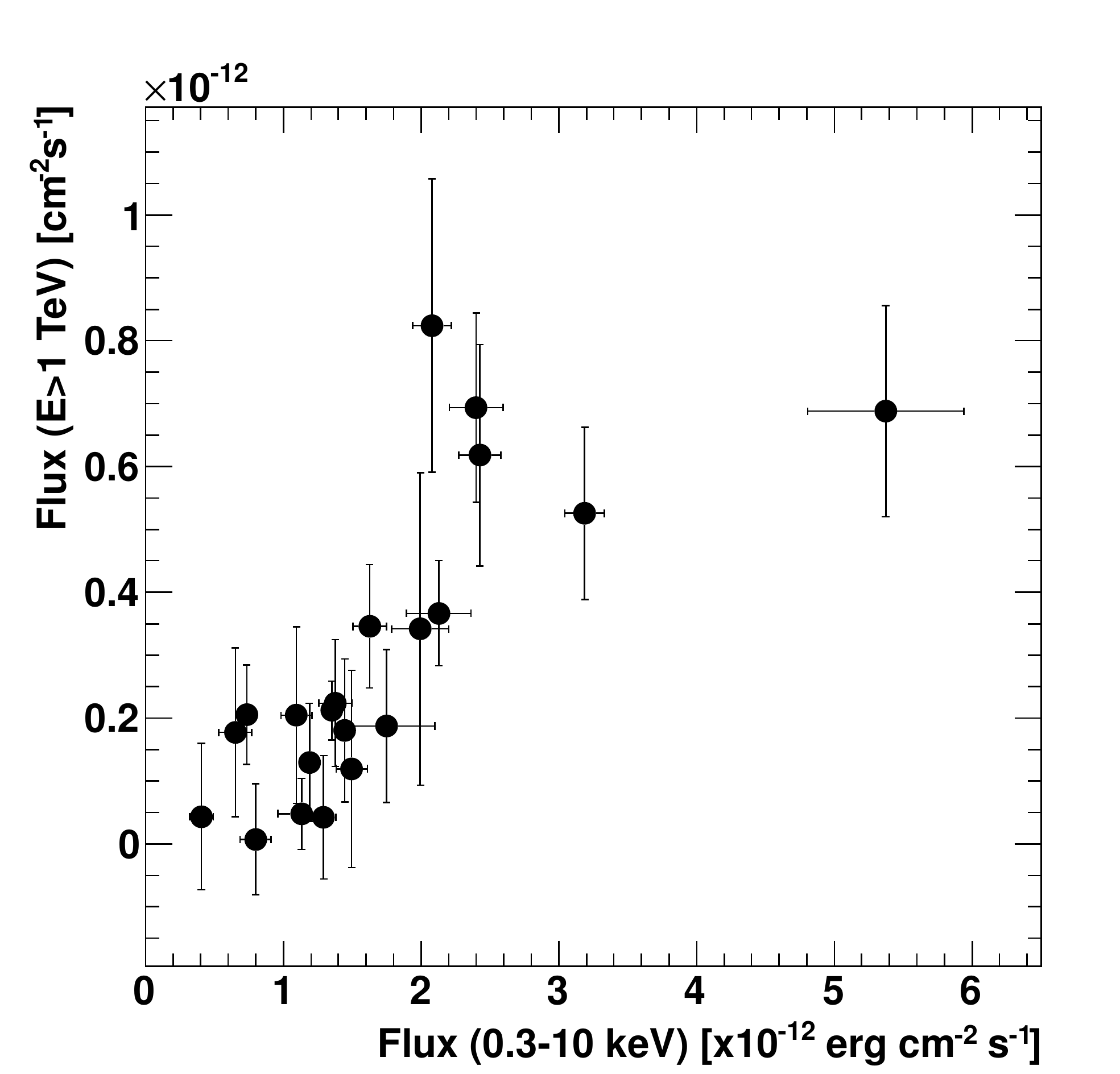}{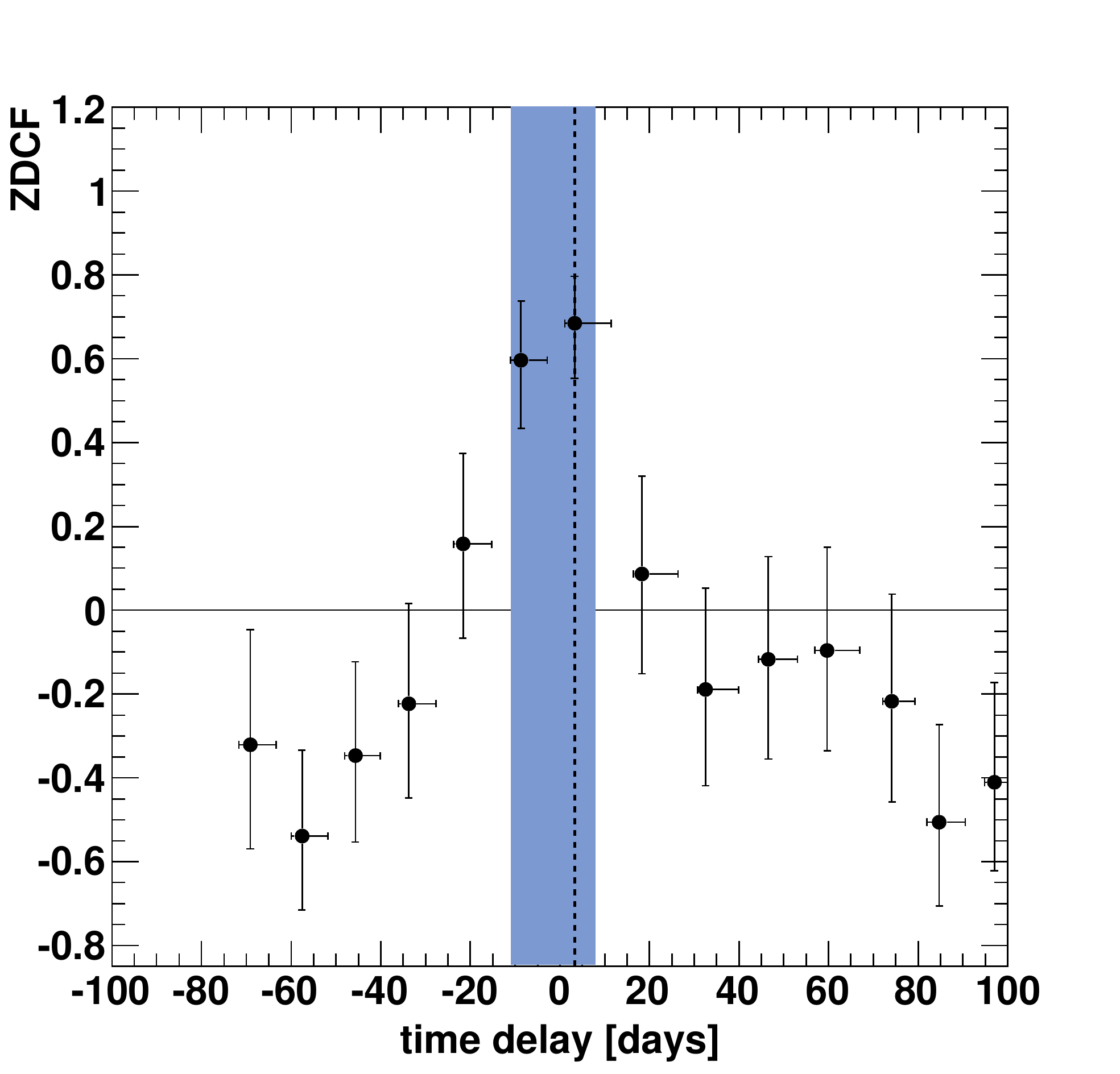}
\caption{\label{fig:ZDCF-VHE-Xray}
Left: Integral $\gamma$-ray fluxes ($>$ 1 TeV) vs  X-ray (0.3--10 keV) fluxes for contemporaneous VHE/X-ray observations. X-ray data were selected
in a $\pm 2.5$ day interval centered on the VHE observing dates. Right: Z-transformed discrete correlation  (Z-DCF) between the
\emph{Swift}-XRT light curve and the combined VERITAS and H.E.S.S. gamma-ray data.  The error bars denote the $1\sigma$ sampling errors
resulting from a MC-based error calculation as described in the text. The dashed line and the blue band indicate the most likely time
lag between X-ray and VHE data of  $3.3^{+8.1}_{-10.8}$ days and the corresponding 68\% fiducial interval.}
\end{figure}

The correlation between gamma-ray and X-ray fluxes for 21 roughly contemporaneous observations is shown in Fig.~\ref{fig:ZDCF-VHE-Xray}. 
X-ray data were selected within a $\pm 2.5$ day interval around the VHE observing dates.  Emission in these energy bands is significantly
correlated (ZDCF/ZDCF$_{\mathrm{error}}$=5.6 at a time lag of zero). The time lag between gamma-ray and X-ray data is consistent with zero 
($\tau_{ZDCF} = +3.3^{+8.1}_{-10.8}$ days).

%
% linear correlation coefficient: 0.7284
% from ZDCF: 5.1 sigma at time lag = 0
%                       5.6 sigma at time lag = -7.7 (most probable)
% correlation analysis with 100,000 MC data sets, see Heike's macro
%     getCorrelationProbability( "VHE-XRAY.dat", 1000000 );
% randomized pairs from VHE and XRT flux, additional randomize inside flux errors, calculate for each set the correlation coefficient.
%

In Fig.~\ref{fig:VTS-ESpec} the differential energy spectra are shown for gamma-ray energies above 200~GeV during the high-flux phases 0.2--0.4 and at phases
0.6--0.9. Figure~\ref{fig:SED} shows the broad-band spectral energy distribution from X-ray to TeV energies.  The shapes of the individual VHE spectra are
consistent with a power-law distribution; see Table \ref{table:EnergySpectrum} for further details.

% $dN/dE = N_{0}(E/1TeV)^{-\Gamma}$ with
%a photon index $\Gamma=2.59\pm0.10_{stat}\pm0.2_{sys}$ and a flux normalization constant  $N_{0}=(0.77\pm0.05_{stat}\pm0.15_{sys}
%)\times 10^{-12}$ cm$^{-2}$ s$^{-1}$ TeV$^{-1}$. 

The differential energy spectrum at VHEs has been measured for three different orbits (2010, 2011 and 2012) and no significant variability in photon index or flux normalization is observed. The spectral results are in agreement with those reported in \cite{Aharonian-2007} and
\cite{Aleksic-2012}. The H.E.S.S. and VERITAS measurements presented here are fully compatible within statistical and systematic uncertainties.
It should be noted that  while the phase ranges for these spectral analyses are similar, the  coverage by observations
inside this phase range is very different for different observation campaigns.  

%The \emph{Swift} spectra are those described in Section \ref{Observations}.

\section{Discussion}
\label{Discussion}
%--------------------------------------------------------------------------------------------------------------------------------------------

The long-term X-ray and TeV observations of HESS~J0632+057 at X-ray energies reported here allow for the first time the modulation of the source
gamma-ray flux to be characterized in a wide orbital phase range, making use of a refined value of the orbital period of the system derived from an
updated X-ray data-set. Below, the implications of the results are briefly discussed, focusing on the phase-folded X-ray and TeV light curves and on
the correlation of the emission observed at both energy bands. The findings are put in the context of current scenarios proposed to explain the
high-energy emission in gamma-ray binaries, and compared to the results obtained for other similar systems.

\subsection{On the X-ray/TeV phase-folded light curves of HESS~J0632+057}

The X-ray light curve of HESS~J0632+057 shows two distinct periods of enhanced emission (see Fig.~\ref{fig:XRT-315d}). The first one,
sharper and higher, appears at orbital phases $\sim 0.3$, which corresponds to about 100 days after periastron passage
\citep{Casares-2012}. The second one is found at phases between $\sim$~0.6--0.9. It appears to be of broader profile with a lower
flux peak, although it is more irregularly sampled in the \emph{Swift}-XRT data set (the orbital period $\sim$315~d and the source
position with respect to the sun made \emph{Swift} unable to cover this phase-range in detail during the previous few cycles). At VHEs, the
source has been repeatedly detected at orbital phases $\sim 0.3$  with the VERITAS, H.E.S.S., and MAGIC observatories. In addition, the
analysis of the H.E.S.S. data at orbital phases in the range 0.6--0.9 reported here (see \S\ref{Results}) has yielded a detection
of the source at TeV energies for the first time in this phase range, in rough coincidence with the secondary bump observed at X-rays.
However, only a few data points characterize the emission at these orbital phases. 
%A likelihood ratio test could not confirm it to constitute a secondary peak in the VHE light-curve at a high significant level, as it is seen in X-rays, so further observations are encouraged to check for %this possibility. 

A double X-ray peak pattern has been observed in other gamma-ray binaries. In LS\,I\,+61\,303, a sharp X-ray peak arises at orbital phases  $\sim0.6$
(periastron is at phase $\sim0.2$),  whilst a broader second peak at orbital phases in the range 0.8--1.0 has been reported, e.g.,~in
\cite{Anderhub-2009} (see however \citealt{Li-2011, Li-2012} and \citealt{Chernyakova-2012} for a study on the long-term evolution of the phase-folded
X-ray light curve of the source). A similar behavior is seen also in the X-ray light curve of the newly-discovered system 1FGL~J1018.6-5856
\citep{Fermi-2012}. In this case, however, the orbital parameters are still lacking, and a correspondence of the position of the two peaks in the
phase-folded light curve and the relative orientation of the compact object and the companion star has not been derived yet. For the well-studied
system LS~5039, which contains an O-type companion star as in 1FGL~J1018.6-5856, both the X-ray and TeV maxima are produced close to the system
inferior conjunction, when the compact object is in front of the star, with no distinct double-peak structure in the X-ray light curve
\citep{Takahashi-2009,Hoffmann-2009}. In the case of PSR~B1259-63, composed of a pulsar and a Be companion star, enhanced X-ray and TeV emission is
found close to periastron. A double X-ray peak has been observed at these orbital phases, but in this case they have been interpreted as the compact
object twice crossing the companion's circumstellar disk \citep{Chernyakova-2009}. No double-peak structure has been claimed yet at TeV energies in the data
collected in the 2004, 2007 and 2011 periastron passages \citep{HESS-Collaboration-2013}. Finally, we note that the peak and dip structure in the
X-ray light curve of HESS~J0632+057 resembles that observed in Eta Carinae, which is thought to be due to the strong interaction of stellar winds as
well as to the geometrical properties of the system orbit \citep{Corcoran-2005}. However, most of the X-ray flux cannot arise from the shocked
stellar wind as the resulting thermal spectrum does not fit well the observed hard X-ray spectrum  \citep{Falcone-2010}.

%%%
% COMMENT ON THE OVERALL X-RAY LIGHTCURVE SHAPE IN LS 5039 AND PSR B1259: NO DOUBLE X-RAY PEAK IN THE FIRST CASE, DOUBLE PEAK BUT ONLY CLOSE TO
% PERIASTRON IN THE SECOND CASE
%%%

The VHE emission observed in known compact gamma-ray binaries can be strongly modulated along the binary orbit due to gamma-ray absorption in the
photon field of the companion star (see, e.g., \citealt{Bottcher-2005, Dubus-2006b} and references therein). In addition, in leptonic
models a relatively high target photon field density is required for gamma-rays to be produced through Inverse Compton (IC) emission processes, whilst
the anisotropy of the companion star photon field further introduces a phase dependence of the spectrum of the upscattered photons (see, e.g., \citealp{Jackson-1972,Dubus-2010}). The detection of VHE emission is therefore subject to the balance between the intrinsic gamma-ray flux and the attenuation factor, which will depend on
the system geometry and therefore on the orbital phase. Considering the orbital parameters recently obtained for HESS~J0632+057 \citep{Casares-2012}, the VHE emission observed in the phase-folded light curve at phases 0.2--0.4 and 0.6--0.9 does not correspond to orbital phases in which the compact object is found close to the Be companion nor close to inferior conjunction where a low opacity to gamma-ray propagation is expected (note however that there are large uncertainties
in the orbital  solution for MWC~148, see \citealp{Casares-2012}).

Other factors may nevertheless contribute to and even dominate the observed modulation in HESS J0632+057. In
particular, fluxes may be regulated by the variability of the underlying particle population emitting at
VHEs. Such variability could come from orbital-dependent adiabatic losses, which may eventually
constrain the maximum energies that particles can attain, or from a discontinuous particle injection,
either in a varying wind-wind shock boundary in a pulsar scenario or due to a phase-dependent accretion
rate in a microquasar model. 
%
%A detailed study considering in detail both the modulation of the intrinsic
%injection of VHE photons and its further attenuation is required to properly characterize the system
%properties at each orbital phase. 
%

It is worth noting that, facing a similar situation in the case of LS~I~+61~303, numerical simulations have shown that a shifted peak in the
high-energy emission light curve may appear in an accretion-based scenario (\citealp{Romero-2007}; see also \citealp{Hayasaki-2005,Orellana-2007}).
A shift of the TeV peak of $\sim 0.3$ orbital phases from periastron is also obtained in a pulsar scenario for the same source by
\citet{Sierpowska-Bartosik-2009} (see also \citealp{Zdziarski-2010}). For both models there is a strong  dependence of the orbital modulation of the
VHE emission on the geometry of the system. The orbital parameters are not known to the required level of accuracy for a more detailed  comparison of
the observed emission pattern with the model predictions. In addition, adiabatic losses could also be responsible for the X-ray and TeV double-peak
profile in the light curve of LS\,I\,+61\,303 (\citealp{Zabalza-2011}; see also \citealp{Takahashi-2009} in the case of LS~5039). The physical
processes leading to this {\it{ad~hoc}} adiabatic loss pattern are however not clear. If the system contains a Be star, they could be related to the
structure of the stellar wind, with two distinct polar and equatorial components (see, e.g., \citealp{Waters-1988}), or to perturbations of the Be
circumstellar disk carried along or affecting orbital phases away from the closest approach during periastron passage. This would affect the emission
properties in both an accretion and a pulsar-wind-based model. In the former case, the total X-ray and VHE fluxes depend linearly on the mass
accretion rate, which depends in turn on the companion's wind density and on the relative velocity with respect to the compact object. If the power
engine is a fast rotating pulsar, instead, the wind profile may also affect the emitter position with respect to the companion star, which would lead
to different emissivities through IC upscattering of the star's photon field. The true nature of the observed modulation of the gamma-ray and X-ray light curves
in HESS~J0632+057 is not yet univocally determined.

\subsection{On the X-ray/TeV correlation in HESS~J0632+057}

\begin{figure}
\plotone{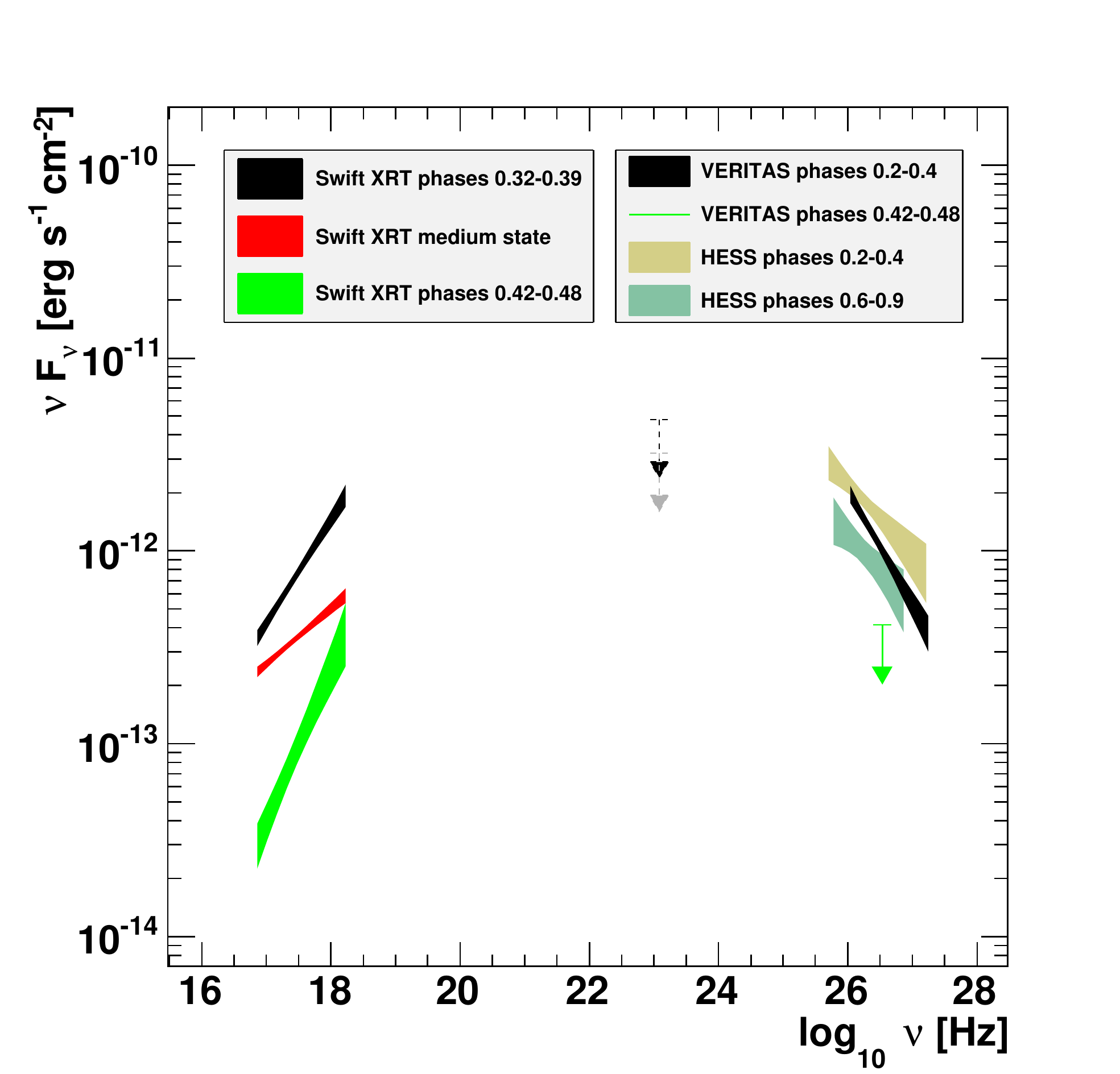}
\caption{\label{fig:SED}
Spectral energy distribution for HESS J0632+057 as measured by
VERITAS, H.E.S.S., \emph{Swift}-XRT, and \emph{Fermi}-LAT \citep{Caliandro-2012}.
The downwards pointing arrow at VHE is an upper flux limit at 95\% confidence level 
for phases 0.42--0.48.
Details for the high, medium and low state X-ray measurements are given in Table \ref{table:SwiftSpectralResults}.
The integral upper limit above 100 MeV measured by \emph{Fermi}-LAT  (confidence level 95\%) has been converted to a differential flux value assuming a spectral index $\Gamma=-2.$ (black dashed arrow)
and $\Gamma=-2.5$ (grey dashed arrow) and is shown at an energy of 500 MeV.
The butterfly plots indicate the 1 sigma statistical errors of the power-law fit to the data points.
}
\end{figure}

The results reported in \S\ref{Results} show a clear correlation between X-rays and TeV gamma-rays. Together with the observed periodicity, this
correlation suggests a causal link between the emission at both energies, for instance produced by processes related to the same population of
accelerated particles. The spectral energy distribution (SED) from X-rays to VHE gamma-rays of HESS~J0632+057 (Fig.~\ref{fig:SED}) reveals a shape
typical of non-thermal high-energy gamma-ray emitters and, in particular, resembles that of known TeV binaries, displaying hard X-ray and soft TeV
spectra. Such spectral shapes can be modeled with relatively simple one-zone leptonic models, as shown, e.g., in \citet{Hinton-2009} and
\citet{Aleksic-2012}, supporting the assumption that X-rays are produced through synchrotron emission of high-energy particles which, in turn,
produce the VHE emission through IC upscattering off the companion's photon field. Note however that, in contrast to other known gamma-ray
binaries, HESS~J0632+057 has not been detected at MeV-GeV energies, despite deep searches using $\sim$~3.5 years of \emph{Fermi}-LAT data
\citep{Caliandro-2012}. This could point to a similar missing correlation between GeV and TeV emission as observed in LS~I~+61~303 or LS~5039
\citep{Hadasch-2012}.

We assume in the following that the particles dominating the emission at the X-ray maximum at phases $\sim$0.3 come indeed from the same population
that is responsible for the TeV emission. We explore here the possibility that a cessation or reduction of the acceleration is the main factor
responsible for the peak-to-dip transition in the X-ray light curve (phases $\sim$0.3--0.4), rather than absorption processes
(\citealp{Falcone-2010,Bongiorno-2011}; see however \citealp{Rea-2011}). Note that this could also imply that the X-ray emission within the dip may
be dominated by a different parent particle population than that seen during the peak. In this transition, a rough characterization of the system
properties can be derived. Particles will lose their energy either through radiative (mainly IC and synchrotron emission) or non-radiative processes
(e.g. adiabatic expansion). In the first case, the ratio of IC over synchrotron losses as a function of the particle energy can be estimated through
$r_{\rm IC/sync} \equiv f_{\rm KN}\,$U$_{\rm rad}/$U$_{\rm mag}$, where $U_{\rm rad}$ and $U_{\rm mag}$ are the photon and magnetic field energy
densities, respectively, and the factor $f_{\rm KN}$  accounts for Klein-Nishina (KN) effects in the IC cross-section  (\citealp{Moderski-2005}; see
also \citealp{Hinton-2009}). $U_{\rm rad}$ can be estimated assuming a distance $d \sim a \approx$~2.38~AU from the emitter to the companion star
($a$ is the semimajor axis of the orbit; \citealp{Casares-2012}), with a radius $R_{\star}= $6.0$~R_{\odot}$ and a temperature  $T_{\rm eff}=
$27.500~K \citep{Aragona-2010}, which yields a peak of the target photon field at $\epsilon_{\rm 0}/m_{\rm e}c^{2} \approx 1.3 \times 10^ {-5}$. For
$U_{\mathrm{mag}}$, \citet{Moldon-2011} reported the detection of radio emission from HESS J0632+057 observed close to the X-ray peak in 2011.  The
authors favor a synchrotron origin for this radio emission, and assuming equipartition of the magnetic field energy with the kinetic energy of the
emitting electrons they derive a value $B \approx 20$ mG. The radio and VHE emission regions could be very different in size and location, however,
so they  may not be characterized by the same magnetic field energy density. Assuming a range of magnetic field values $B = 10, 20$~and~50~mG,  the
ratio  $r_{\rm IC/sync} = 1$ would correspond to electron energies $E_{\rm e} \approx$~8.7,~3.4, and 1.0~TeV, respectively. For lower and higher
energies, IC and synchrotron processes would correspondingly dominate the total radiative losses. Note however that the upper limits at GeV energies
reported in \citet{Caliandro-2012}, and the lack of information at hard X-rays, prevent a more accurate evaluation of the energy output channeled
through both radiation mechanisms.

Further constraints can be obtained from the X-ray phase-folded light curve. If the {\it{Swift}}-XRT flux
has a synchrotron origin, electrons with energies
$E_{\rm e} \approx 0.285 \, (E_{\rm sync}/5.4$~keV)$^{1/2} (B/1\mathrm{G})^{-1/2}$~TeV are required
($E_{\rm sync}$ is the characteristic synchrotron energy of the emitted photon). For the range of
magnetic field values $B = 10, 20$,~and~50~mG, $E_{\rm e} \sim 4.5, 3.2$, and 2.0~TeV, respectively, and
from the considerations above, both synchrotron and IC processes would contribute significantly to the
radiative particle cooling (see, e.g., \citealp{Skilton-2009}). The total time scale for the radiative losses, $t_{\rm rad} = (t_{\rm
sync}^{-1} + t_{\rm IC-KN}^{-1})^{-1}$ would range between $\sim$~1.5~d and $\sim$~2~d. This
time scale is roughly comparable to that in which the X-ray flux is observed to decrease
by a noticeable factor in the transition from the X-ray maximum to the X-ray dip in Fig.
\ref{fig:VHE-XRT} ($t_{\rm peak-dip} \sim 0.01$ phases corresponding to $3.3$~d using $P = 315$~d).

Regarding adiabatic losses, \citet{Moldon-2011} report also on the detection of extended radio emission
from HESS J0632+057 seen 30 days after the X-ray peak in January/February 2011\footnote{\footnotesize{Note however that the flux
obtained for this extended radio emission is below the RMS level of that derived for the point-like
source emission obtained close to the X-ray maximum.}}. The total extension of this emission was of the
order of 50 mas, which translates to about $l_{\rm ext}\sim$75 AU when a distance of $\sim 1.5$~kpc
to the source is assumed. If material ejected from the vicinity of the compact object expanded to reach
$l_{\rm ext}$ in $\lesssim 30$~days, an expansion velocity $v_{\rm exp} \gtrsim 1.2 \times
10^{8}$~cm~s$^{-1}$ would imply an adiabatic cooling time $t_{\rm ad}\sim l_{\rm ext}/v_{\rm exp}
\lesssim 108$~d. Therefore, radiative cooling may have dominated the total losses during the X-ray
peak-to-dip transition, unless the ejected material expanded at a high velocity within the 30~d lag
between the two radio observations, $v_{\rm exp} \sim 0.25\,c$, in which case $t_{\rm ad} \approx
t_{\rm rad}$.

These crude estimates should be seen more as illustrative than as a detailed description of the true physical processes leading to the observed
high-energy emission in HESS~J0632+057, and alternative scenarios may be considered. On the one hand, electrons could produce X-rays through IC instead
of synchrotron processes, reaching also the gamma-ray domain. In that case, however, the wide range of electron energies required would imply very
different cooling time scales, at odds with the tight correlation observed in the X-ray/TeV light curve. Furthermore, the lack of GeV emission
\citep{Caliandro-2012}, would be difficult to justify in such scenario. On the other hand, VHE emission could be produced through hadronic
interactions of protons accelerated close to the compact object against ions present in the companion's wind and/or circumstellar disk. In parallel,
secondary electrons/positrons would then be produced through pion decay in the same proton-proton interactions that would initiate the gamma-ray
fluxes, leading to an X-ray energy flux at a level $\sim 1/2$ of that produced at gamma-rays (see, e.g., \citealp{Kelner-2006}). This is not
observed, however, during the X-ray/TeV peak at phase $\sim0.3$ (see Fig.~\ref{fig:SED}), in which a similar luminosity is radiated at both X-rays
and TeV energies. Note also that there is no evidence of a cutoff of the X-ray spectrum, which is relatively hard with a spectral index of $1.46 \pm
0.06$, and which may even extend into the hard X-ray domain, enhancing the total X-ray luminosity of the source. 

%   
% Seemingly, the Fermi non-detection of HESS J0632+057 implies the existence of a turnover of the gamma-ray spectrum at energies $\sim 1 -
% 100$~GeV (see also \citet{Aleksic-2012}). This would prevent from a substantial increase of the total gamma-ray luminosity from the source when %extending the spectrum down into the HE regime,
% and the ratio of X-ray to gamma-ray fluxes would not be gerdied. If the hadronic interactions would take place at distances larger than the typical binary-size length-scale, instead, then
% the condition of a copious secondary e$^{\pm}$ production would be relaxed at some extent (see, e.g., \cite{Khangulyan-2008, Bosch-Ramon-2008}). In such scenario, it would be however difficult
% again to reconcile the very similar cooling time-scales, as related to independent populations which may in turn be located at different regions in the system. 
%

Finally,  properties of the medium like the radiation and matter density fields, rather than the
intrinsic properties of the emitter, could be responsible for the simultaneous modulation that
shapes the observed fluxes at X-ray and gamma-ray energies in a periodic way. However, it should be noted that VHE
radiation is affected mainly by the interaction with the companion star photon field. Conversely, X-ray
fluxes are mainly reprocessed through interactions with the ambient matter. In this regard, an
enhancement of the local matter density might be expected during the X-ray light curve minima, which has
not been observed (see, e.g., \citealp{Bongiorno-2011, Rea-2011}). This, together with the tight X-ray/TeV
correlation implying a similar modulation of the fluxes at both energy bands, favors a scenario in which
the variability arises from a modulation of a common underlying emitting-particle population.

A correlation of X-ray and TeV emission has also been observed in other gamma-ray binaries. In particular, correlated X-ray/VHE
emission has been reported in the case of LS~I~+61~303 through simultaneous MAGIC, XMM-Newton, and \emph{Swift}-XRT observations in a
multiwavelength campaign conducted in 2007 \citep{Anderhub-2009}. The correlation result was not apparent however in later observations
of the source \citep{Acciari-2011}, and a change in the source/medium properties has also been observed in recent gamma-ray
observations \citep{Acciari-2011,Aleksic-2012, Hadasch-2012}, which display strong deviations of the source phase-folded flux profiles
as compared to older data. The processes leading to such transitions in the modulation of the light curve are not clear, although a relation to
the superorbital variability of the companion star seen at lower radio and X-ray energies has been suggested
(\citealp{Gregory-2002}; see also \citealp{Li-2012, Chernyakova-2012}). As compared to LS~I~+61~303,
the emission from HESS~J0632+057 seems to be steadier, with a remarkably lower orbit-to-orbit variability and only small
deviations from the main, long-term pattern observed for more than five years. Further observations are
required to assess whether a superorbital modulation is also present in HESS~J0632+057.

%%%% GM to here: only LS I +61 303 is discussed, without a lot of references to LS I +61 303

%\subsubsection{Spectral energy distribution}

%The spectral energy distribution (SED) from X-rays to VHE gamma-rays of HESS J0632+057 (Fig.~\ref{fig:SED}) reveals a  shape typical for  non-thermal
%high-energy gamma-ray emitters. Such spectral shape can be modeled with very simple one-zone leptonic models, as shown e.g. in \citet{Hinton-2009} and
%\citet{Aleksic-2012}, supporting the assumption that the observed emission at X-ray and gamma-ray energies may be produced by the same particle
%population. However, these models should be seen as more illustrative than as a detailed description of the true physical processes leading to the
%observed high-energy emission in HESS J0632+057. 

%\citet{Zabalza-2012} consider for example acceleration of particles at two different sites: in the head-on
%collision zones of the winds and  in the termination shock caused by Coriolis forces away from the dense photon field of the bright star.

\section{Final remarks}

HESS~J0632+057 together with LS I +61 303 and PSR B1259-63 is one of the three gamma-ray binaries known to contain a Be companion star.  Common
processes leading to the production of non-thermal emission from radio to VHE gamma rays may explain the broad-band energy distribution in all of
them. However, differences in their orbital parameters and the nature of the still-unknown power sources in LS I +61 303 and HESS~J0632+057, may
ultimately define their individual observational properties, including the phase-folded patterns observed in each case. Detailed models with a realistic description of the  geometry of the orbit, the interaction of the stellar wind with the wind or jet of the compact object, and the distribution of photon and matter fields, are necessary to get a deeper understanding of the system and its orbital variability. 

%Finally, we want to remark the limitations of the considerations above given the known inhomogeneities of winds from Be stars (e.g. \citet{Perucho-2012}) and the expected turbulences present in the wind-wind interaction zones of the binary system.

Due to their variable and relatively well-constrained environment, the characterization of the high-energy behavior of gamma-ray binaries
has become an important research field in recent years. New candidates have been discovered (e.g.~1FGL~J1018.6-5856; \citealp{Fermi-2012,
Abramowski-2012}) and unexpected features are being observed in some of the known sources as, e.g., in PSR~B1259-63 (\citealp{Abdo-2011})
and LSI~+61~303 (\citealt{Acciari-2011}). Furthermore, detailed numerical simulations are being run
(e.g. \citealp{Romero-2007, Sierpowska-Bartosik-2008, Perucho-2010, Perucho-2012, Bosch-Ramon-2012}) 
%(e.g. \citealp{Romero-2007, Perucho-2010, Perucho-2012, Bosch-Ramon-2012}) 
as well as new scenarios are being proposed to explain them (see, e.g.,
\citealp{Khangulyan-2012}; see also \citealp{Torres-2012, Zabalza-2013, Bednarek-2013}). These and future studies, together with the improved
capabilities of next-generation VHE observatories, may provide new clues to unveil the physics behind gamma-ray binaries.

\acknowledgments
%--------------------------------------------------------------------------------------------------------------------------------------------

This research is supported by grants from the U.S. Department of Energy Office of Science, the U.S. National Science Foundation and the Smithsonian Institution, by NSERC in Canada, by Science Foundation Ireland (SFI 10/RFP/AST2748) and by STFC in the U.K. We acknowledge the excellent work of the technical support staff at the Fred Lawrence Whipple Observatory and at the collaborating institutions in the construction and operation of the instrument.
GM acknowledges support through the Young
Investigators Program of the Helmholtz Association and the Helmholtz Alliance for Astroparticle Physics.

\noindent The support of the Namibian authorities and of the University of Namibia in facilitating the construction and operation of H.E.S.S. is gratefully acknowledged, as is the support by
the German Ministry for Education and Research (BMBF), the Max Planck Society, the French Ministry for Research, the CNRS-IN2P3 and the Astroparticle Interdisciplinary Programme of the CNRS,
the U.K. Particle Physics and Astronomy Research Council (PPARC), the IPNP of the Charles University, the South African Department of Science and Technology and National Research Foundation,
and by the University of Namibia. We appreciate the excellent work of the technical support staff in Berlin, Durham, Hamburg, Heidelberg, Palaiseau, Paris, Saclay, and in Namibia in the
construction and operation of the  H.E.S.S. equipment.

{\it Facilities:} \facility{VERITAS}, \facility{H.E.S.S.}, \facility{Swift}

%%%%%%%%%%%%%%%%%%%%%%%%%%%%%%%
% (GM) table checked 2012/09/15
%%

\begin{deluxetable}{cccccccccc}
\tabletypesize{\scriptsize}
\centering
\tablecolumns{9}
\tablewidth{0pt}
\tablecaption{H.E.S.S. and VERITAS analysis results for energies E $> 1$ TeV.
\label{table:ObservationsAndResults}}
\tablehead{
\colhead{MJD} &
\colhead{Mean} &
\colhead{Observatory} &
\colhead{Observation} &
\colhead{Mean} &
\colhead{On} &
\colhead{Off} &
\colhead{$\alpha$\tablenotemark{b}} &
\colhead{Significance\tablenotemark{c}} &
\colhead{Flux} \\
\colhead{Range} &
\colhead{phase\tablenotemark{a} } &
\colhead{} &
\colhead{Time} &
\colhead{Elevation} &
\colhead{Events} &
\colhead{Events} &
\colhead{} &
\colhead{($\sigma$)} &
\colhead{(Upper Flux Limit\tablenotemark{c})} \\
\colhead{} &
\colhead{} &
\colhead{} &
\colhead{(minutes)} &
\colhead{(deg)} &
\colhead{} &
\colhead{} &
\colhead{} &
\colhead{(pre-trial)} &
\colhead{($10^{-13}$ cm$^{-2}$ s$^{-1})$} 
}
\startdata

\hline
54089 - 54125 & 0.62 & VERITAS & 579 & 61.3 & 12 & 153 & 0.06 & 1.1 & 1.4 $\pm$ 1.5 $( < 6.1)$ \\
54830 - 54834 & 0.92 & VERITAS & 561 & 62.6 & 7 & 180 & 0.05 & -0.9 &  -0.9 $\pm$ 0.9 $(< 2.3)$ \\
54856 - 54861 & 0.01 & VERITAS & 721 & 61.9 & 12 & 197 & 0.05 & 0.4 &  0.4 $\pm$ 1.0  $(< 3.6)$ \\
55122 - 55133 & 0.86 & VERITAS & 491 & 62.6 & 9 & 78 & 0.05 & 2.0 &  1.8 $\pm$ 1.1  $(< 4.5)$ \\
55235 - 55247 & 0.22 & VERITAS & 925 & 61.3 & 25 & 136 & 0.05 & 4.9 & 3.5 $\pm$ 1.0 \\
55259 - 55276 & 0.30 & VERITAS & 309 & 59.5 & 18 & 55 & 0.05 & 5.6 & 8.2 $\pm$ 2.3 \\
55544 - 55564 & 0.21 & VERITAS & 229 & 62.7 & 4 & 46 & 0.05 & 0.9 &   1.2 $\pm$ 1.6 $(< 7.1)$ \\
55571 - 55572 & 0.27 & VERITAS & 140 & 61.7 & 4 & 23 & 0.05 & 1.9 & 3.4 $\pm$ 2.5 $(< 12.8)$ \\
55585 - 55599 & 0.33 & VERITAS & 639 & 58.6 & 25 & 106 & 0.05 & 5.7 & 5.3 $\pm$ 1.4 \\
55600 - 55602 & 0.36 & VERITAS & 541 & 58.9 & 27 & 104 & 0.05 & 6.2 & 6.9 $\pm$ 1.7 \\
55614 - 55622 & 0.42 & VERITAS & 643 & 58.7 & 14 & 105 & 0.05 & 2.9 & 2.2 $\pm$ 1.0 $(< 5.5)$ \\
55624 - 55630 & 0.45 & VERITAS & 342 & 57.8 & 5 & 77 & 0.05 & 0.4 & 0.4 $\pm$ 1.2 $(<4.6)$ \\
55643 - 55656 & 0.52 & VERITAS & 468 & 53.7 & 14 & 158 & 0.05 & 1.7 &  2.0 $\pm$ 1.4  $(< 6.5)$ \\
55891 - 55901 & 0.30 & VERITAS & 454 & 61.0 & 21 & 90 & 0.05 & 5.2 & 6.2 $\pm$ 1.8 \\
55916 - 55920 & 0.37 & VERITAS & 632 & 59.8 & 31 & 98 & 0.05 & 7.3 & 6.9 $\pm$ 1.5 \\
55921 - 55927 & 0.39 & VERITAS & 419 & 62.4 & 8 & 67 & 0.05 & 1.9 &  1.9 $\pm$ 1.2 $(< 6.0)$ \\
55940 - 55949 & 0.45 & VERITAS & 295 & 62.4 & 2 & 35 & 0.05 & 0.1 & 0.1 $\pm$ 0.9   $(< 3.8)$ \\
55951 - 55955 & 0.48 & VERITAS & 256 & 62.7 & 4 & 25 & 0.05 & 1.8 & 1.8 $\pm$ 1.3   $(< 6.8)$ \\
\hline
53087 - 53088 & 0.38  &  HESS &   77.7  &  43.9  &   13  &   185  &  0.05  &  1.1  & 1.3 $\pm$ 1.4  $(< 4.2)$  \\ 
53353 - 53356 & 0.23  &  HESS &  290.6  &  54.3  &  110  &  1209  &  0.05  &  6.0  & 4.0 $\pm$ 0.8  \\  
53685 - 53716 & 0.32  &  HESS &  324.7  &  53.4  &  113  &  1175  &  0.04  &  7.2  & 4.9 $\pm$ 0.9  \\
53823 - 53823 & 0.71  &  HESS &   79.1  &  47.5  &   22  &   328  &  0.03  &  3.0  & 4.7 $\pm$ 1.9  \\
54117 - 54118 & 0.65  &  HESS &  254.9  &  58.7  &   72  &   933  &  0.05  &  3.1  & 2.0 $\pm$ 0.7  \\ 
54169 - 54170 & 0.81  &  HESS &   54.3  &  59.7  &    5  &   141  &  0.05  &  -0.8 & -0.9 $\pm$ 0.09 $(< 1.4)$	   \\  
54414 - 54426 & 0.61  &  HESS &  156.4  &  59.5  &   50  &   538  &  0.05  &  3.7  & 3.3 $\pm$ 1.0  \\
54467 - 54475 & 0.77  &  HESS &  217.1  &  58.4  &   69  &   644  &  0.05  &  5.2  & 3.9 $\pm$ 0.9  \\
54859 - 54910 & 0.08  &  HESS &  161.6  &  52.1  &   26  &   368  &  0.05  &  1.5  & 1.3 $\pm$ 0.09  $(< 3.3)$	   \\ 
55121 - 55157 & 0.89  &  HESS &  643.0  &  59.7  &  140  &  1664  &  0.05  &  5.2  & 2.1 $\pm$ 0.5  \\  
55178 - 55185 & 0.02  &  HESS &  437.5  &  55.1  &   51  &   869  &  0.05  &  0.8  & 0.5 $\pm$0.6 $(< 1.6)$	   \\
55895 - 55898 & 0.29  &  HESS &  230.7  &  59.4  &   87  &   885  &  0.05  &  5.2  & 3.6 $\pm$ 0.8  \\
55931 - 55951 &	0.44  &  HESS &  233.2  &  52.5  &   53  &   673  &  0.05  &  2.9  & 2.1 $\pm$ 0.8 $(< 3.7)$	   \\

%53353 - 53353 & 0.23 & HESS & 283.8 & 54.2 & 22 & 252 & 0.02 & 5.3 & 6.4 $\pm$ 1.8 \\  
%53685 - 53716 & 0.33 & HESS & 307.4 & 53.3 &  7 & 183 & 0.02 & 1.4 & $<$ 3.9 \\
%53823 - 53823 & 0.71 & HESS &  25.8 & 47.1 &  3 &  25 & 0.02 & 2.3 & $<$ 21.1  \\ 
%54117 - 54118 & 0.65 & HESS & 258.6 & 58.8 &  8 & 156 & 0.02 & 2.2 & $<$ 5.9   \\  
%54169 - 54170 & 0.82 & HESS &  51.6 & 59.6 &  1 &  23 & 0.02 & 0.7 & $<$ 14.4  \\
%54414 - 54426 & 0.61 & HESS & 152.4 & 59.5 &  4 &  66 & 0.02 & 1.8 & $<$ 5.6   \\
%54467 - 54475 & 0.78 & HESS & 132.4 & 58.2 &  8 &  97 & 0.02 & 3.1 & 4.7 $\pm$ 2.2 \\    
%54859 - 54910 & 0.09 & HESS & 130.8 & 54.1 &  2 & 114 & 0.02 & 0.3 & $<$ 4.2   \\
%55121 - 55157 & 0.89 & HESS & 628.3 & 59.5 & 12 & 313 & 0.02 & 1.9 & $<$ 3.1   \\
%55178 - 55184 & 0.03 & HESS & 407.4 & 54.8 &  2 & 244 & 0.02 & 1.6 & $<$ 0.5   \\
%55895 - 55898 & 0.29 & HESS & 226.8 & 59.7 & 11 & 127 & 0.02 & 3.8 & 4.4 $\pm$ 1.7  \\
%55931 - 55951 & 0.44 & HESS & 232.8 & 52.8 &  7 & 191 & 0.02 & 1.4 & $<$ 5.2   \\

\enddata
\tablenotetext{a}{Phases are calculated using an orbital period of 315 d and MJD$_{0}$=54857.}
\tablenotetext{b}{$\alpha$ denotes the ratio between the area used for the determination of on and off counts.}
\tablenotetext{c}{Significances are calculated using equation (17) from \cite{Li-1983}.}
\tablenotetext{d}{
Errors on fluxes are $1\sigma$ statistical uncertainties.
Upper limits (E$>1$ TeV)
are given in brackets at 99\% confidence level 
(after \citealt{Rolke-2001}) for periods with a significance lower than $3 \sigma$.
}\end{deluxetable}

%%%%%%%%%%%%%%%%%%%%%%%%%%%%%%%%%%%%%%%%%

\begin{deluxetable}{ccccc}
\centering
\tablecolumns{5}
\tablewidth{0pt}
\tablecaption{Outcome of the spectral analysis of VHE photons.
The table lists the  results of the  power-law fits to the differential energy spectra, see Fig. \ref{fig:VTS-ESpec}.
The H.E.S.S.~and MAGIC results from the literature are taken over a  phase range 0.2--0.5, but with
different observational coverage of the light curve.
Errors are $1\sigma$ statistical errors only. 
The systematic error on the flux constant is typically 20\% and on the spectral index $\approx 0.1$.
\label{table:EnergySpectrum}}
\tablehead{
\colhead{Year} &
\colhead{Orbital} &
\colhead{flux normalization constant ($\times 10^{-13}$)} &
\colhead{photon index} &
\colhead{$\chi^2/N$} \\
\colhead{} &
\colhead{phase} &
\colhead{at 1 TeV [cm$^{-2}$s$^{-1}$TeV$^{-1}$]} & 
\colhead{} &
\colhead{} 
}
\startdata
VERITAS 2010 & 0.2--0.4 & $(6.4\pm1.0) $ & $2.2\pm0.4$ & $1.7/3$ \\
VERITAS 2011 & 0.2--0.4 & $(11.0\pm1.1) $ & $2.5\pm0.2$ & $4.2/4$ \\
VERITAS 2012 & 0.2--0.4 & $(6.4\pm0.8) $ & $2.3\pm0.2$ & $5.8/6$ \\
\hline
H.E.S.S. 2004-2012 & 0.2--0.4 & $(5.7\pm 0.7) $ & $2.3\pm 0.2$ & $32.0/31$ \\
VERITAS 2010-2012 & 0.2--0.4 &  $(7.7\pm0.5) $ & $2.6\pm0.1$ & $6.0/6$ \\
H.E.S.S. 2004-2012 & 0.6--0.9 & $(3.9\pm0.7) $ & $2.4\pm0.2$ & $44.0/27$ \\
\hline \hline
H.E.S.S. 2004/2005\tablenotemark{a} & & $(9.1\pm1.7) $ & $2.53\pm0.6$ & - \\
MAGIC 2012\tablenotemark{b}   & & $(12\pm0.3) $ & $2.6\pm0.3$ &  -
\enddata
\tablenotetext{a}{\citet{Aharonian-2007}}
\tablenotetext{b}{\citet{Aleksic-2012}}
\end{deluxetable}

%%%%%%%%%%%%%%%%%%%%%%%%%%%%%%%%%%%%%%%%%
%
% Swift spectral states
%
%High State:
%---------
%0.3-10 keV flux: 3.71e-12 (+0.12e-12, 0.09e-12)  ergs cm^-2 s^-1
%photon index: 1.46 (+0.06, -0.05)
%norm. const.: 4.25e-4 (+0.25e-4, -0.24e-4)  photons cm^-2 s^-1 keV^-1
%nH: 3.81e21 (frozen to fit value for complete data set)
%chi^2/dof: 80.3/78
%
%Medium State:
%------------
%0.3-10 keV flux: 1.43e-12 (+0.03e-12, -0.03e-12) ergs cm^-2 s^-1
%photon index: 1.71 (+0.04, -0.04)
%norm. const.: 2.09e-4 (+0.08e-4, -0.08e-4) photons cm^-2 s^-1 keV^-1
%nH: 3.81e21
%chi^2/dof: 224.0/168
%
%Low State:
%--------
%0.3-10 keV flux: 6.05e-13 (+0.28e-13, -0.50e-13) ergs cm^-2 s^-1
%photon index: 1.19 (+0.16, -0.16)
%norm. const.: 5.06e-5 (+0.90, -0.83) photons cm^-2 s^-1 keV^-1
%nH: 3.81e21
%chi^2/dof: 6.01/14
%
\begin{deluxetable}{ccccc}
\centering
\tablecolumns{5}
\tablewidth{0pt}
\tablecaption{Outcome of the spectral analysis of the \emph{Swift} X-ray data.
Parameters are derived from fitting an absorbed power law with a fixed 
absorption coefficient N$_{\mathrm{H}}~=~3.81\times~10^{21}$~cm$^{-2}$.
The spectral fit on medium flux states is added for illustration. Data for this fit are 
selected according to their absolute flux values and not in a specific phase range.
\label{table:SwiftSpectralResults}}
\tablehead{
\colhead{flux state} &
\colhead{phase range}&
\colhead{flux normalization constant ($\times 10^{-4})$} &
\colhead{photon index} &
\colhead{$\chi^2/N$} \\
\colhead{} &
\colhead{} &
\colhead{[cm$^{-2}$s$^{-1}$keV$^{-1}$]} & 
\colhead{} &
\colhead{} 
}
\startdata
high & 0.32--0.39 & $(4.25\pm0.24)$ & $1.46\pm0.06$ & $80/78$ \\
medium & - & $(2.09\pm0.08)$ & $1.71\pm0.04$ & $224/168$ \\
low & 0.42--0.48 & $(0.50\pm0.01) $ & $1.19\pm0.16$ & $ 6/14$ 
\enddata
\end{deluxetable}

%%%%%%%%%%%%%%%%%%%%%%%%%%%%%%%%%%%%%%%%%%%%%%%%%%%%%%%%%%%%%%%%%%%%%%%
%%%%%%%%%%%%%%%%%%%%%%%%%%%%%%%%%%%%%%%%%%%%%%%%%%%%%%%%%%%%%%%%%%%%%%%
% BinaryAnalysis.C
% plotLightCurvesMJD( 54800 );
% data/X-ray/Swift/lightcurve_XRT_AF_20120216.dat
% 
% plotLightCurvesMJD_VHE_Xrays(true) or false
% data/HESS/HESS_lc_dates_non_merged_20120917.dat
%

%%%

%%%%%%%%%%%%%%%%%%%%%%%%%%%%%%%%%%%%%%%%%%%%%%%%%%%%%%%%%%%%%%%%%%%%%%%
%%%%%%%%%%%%%%%%%%%%%%%%%%%%%%%%%%%%%%%%%%%%%%%%%%%%%%%%%%%%%%%%%%%%%%%
% see analysis/ZDCF/AutoCorrelation/README for results
% same data as in data/X-ray/Swift/lightcurve_XRT_AF_20120216.dat
%%%

%%%%%%%%%%%%%%%%%%%%%%%%%%%%%%%%%%%%%%%%%%%%%%%%%%%%%%%%%%%%%%%%%%%%%%%
%%%%%%%%%%%%%%%%%%%%%%%%%%%%%%%%%%%%%%%%%%%%%%%%%%%%%%%%%%%%%%%%%%%%%%%
% XRT phase folded light-curve
% .L BinaryAnalysis.C 
% plotLightCurvesPhaseFolded_VHE_Xrays( "data/xrt_0632_ObservingPeriods_315d.dat");

%%%%%%%%%%%%%%%%%%%%%%%%%%%%%%%%%%%%%%%%%%%%%%%%%%%%%%%%%%%%%%%%%%%%%%%%
% VHE phase folded light-curve
% .L BinaryAnalysis.C 
% plotLightCurvesPhaseFolded_VHE_Xrays();
%

%%%%%%%%%%%%%%%%%%%%%%%%%%%%%%%%%%%%%%%%%%%%%%%%%%%%%%%%%%%%%%%%%%%%%%%%
% VHE phase folded light-curve for +1 sigma and -1 sigma
% .L BinaryAnalysis.C 
% plotLightCurvesPhaseFolded_VHE_Xrays();  // with changes in the code
%

%%%%%%%%%%%%%%%%%%%%%%%%%%%%%%%%%%%%%%%%%%%%%%%%%%%%%%%%%%%%%%%%%%%%%%%%
% VHE / XRT correlation plot
% .L $EVNDISPSYS/lib/libVAnaSum.so 
% .L BinaryAnalysis.C 
% plotLightCurvesPhaseFolded_VHE_Xrays(); 
%
%%%%%%%%%%%%%%%%%%%%%%%%%%%%%%%%%%%%%%%%%%%%%%%%%%%%%%%%%%%%%%%%%%%%%%%%
% VHE / XRT ZDCF analysis
% see  see analysis/ZDCF/Correlation/README for details and results
% data files HESSJ0632_lightcurve_flux_16Feb2012.dcfinput.txt and VTS_HESS_20120917.dcfinput.txt
% 
% 5.4 sigma from ZDCF/error
%

%%%%%%%%%%%%%%%%%%%%%%%%%%%%%%%%%%%%%%%%%%%%%%%%%%%%%%%%%%%%%%%%%%%%%%%%
% VTS spectral reconstruction
%
% energy spectra for periods phase 0.2-0.4: this includes all significant VTS VHE points and the X-ray maximum.
%
% energy spectrum in HESS 2007 paper: phase range from 0.23 to 0.38 (2004+2005 data)
% energy spectrum in MAGIC 2012 paper: phase range from 0.36 to 0.4 (limited coverage)
%
% .L BinaryAnalysis.C 
% plotEnergySpectra();

%%%%%%%%%%%%%%%%%%%%%%%%%%%%%%%%%%%%%%%%%%%%%%%%%%%%%%%%%%%%%%%%%%%%%%%%
% SED
%
% SED for VERITAS / LAT / Swift data (HESS data still to come)
%
%

%\begin{figure}
%\plotone{plots/SEDpRadio.pdf}
%\caption{\label{fig:SED-Radio}
%This one?
%}
%\end{figure}

\end{document}